\documentclass[onecolumn, fleqn,usenatbib]{mnras}

\usepackage[T1]{fontenc}
\usepackage{ae,aecompl}

\usepackage{graphicx}	
\usepackage{amsmath}	
\usepackage{amssymb}	

\title[Energy constraints in the IllustrisTNG]{Mesoscopic Energy Ranking Constraints in the IllustrisTNG Simulations}

\author[Dantas, C. C.]{
Christine C. Dantas$^{1}$\thanks{E-mail: christine.dantas@inpe.br}
\\
$^{1}$Divis\~ao de Astrof\'{\i}sica (INPE-MCTI), 
S\~ao Jos\'e dos Campos, 12227-010, SP, Brazil 
}

\date{Accepted XXX. Received \today; in original form 22 January 2022}

\pubyear{2022}

\begin{document}
\label{firstpage}
\pagerange{\pageref{firstpage}--\pageref{lastpage}}
\maketitle

\begin{abstract}
We revisited the problem of mixing in a gravitational N-body system from the point of view of the ordering of coarse-grained cells in the one-particle energy space, here denoted {\it energy ranking preservation} (ERP). This effect has been noted for some time in simulations, although individual particle energies and their phase-space variables mix considerably. The present investigation aimed to map ERP in terms of parameters involving the collective range in which it is effective, as well as in terms of global and historical characterisations of gravitational systems evolving towards equilibrium. We examined a subset of the IllustrisTNG cosmological magnetohydrodynamical simulations (TNG50-4 and TNG100-3), considering both their full and dark-only versions. For each simulation, we selected the $20$ most massive haloes at redshift $z=0$, tracing their ERP fractions back at selected redshift markers ({\tt z} $= \{1.0, 5.0, 10.0 \}$), and for a coarse-graining set ranging from $5$ to $30$ energy bins.  At the redshift marker {\tt z} $= 1$, we found high ERP fractions (above $\sim 80 \%$) in both simulations, regardless of the coarse-graining level. The {\it decline} in ERP fractions with redshift was roughly a function of mass and fractional mass increase in the analysed TNG50-4 haloes, but not in the TNG100-3 ones, indicating a possible relative  susceptibility of the ERP effect to mass accretion for haloes less massive than $\sim 10^{14} ~ M_{\odot}$. We confirmed earlier indications in the literature concerning a possible ``mesoscopic'' constraint operative in a time span of at least several Gyr. 
\end{abstract}




\section{Introduction}
\label{INTRO}

Large-scale structures are probably the result of gravitational amplification of initially small density fluctuations, which grew hierarchically in an expanding cosmological background \citep{Spr06}. State of the art magnetohydrodynamical cosmological simulations, following the $\Lambda$CDM model and including semi-analytic approaches, reproduce the observable Universe in considerable detail and in agreement with the hierarchical instability proposition \citep{Vog20}.

Yet, there are significant open questions regarding the processes leading to relaxation and equilibrium of galaxies and clusters of galaxies. These questions are fundamentally related to the nature of the N-body gravitational problem, as a leading-order catalyst of the dynamical and structural characteristics of those systems. In this regard, the so-called ``violent relaxation'' (VR, i.e., a rapidly varying potential process) has been proposed as an efficient mechanism for redistributing or mixing particles in phase-space \citep{Kin62,Hen64,Lyn67}, so that a nearly steady state could be reached after just a few crossing times of the system. 

However, VR is still somewhat a qualitative hypothesis, with unknown elements yet to be understood for a complete theory, so that currently formalisation attempts are being pursued in several fronts \citep[and references therein]{Bar22}. For instance, it has been noted since the 80's that, after the  occurrence VR, not all dynamically accessible phase-space cells ended up being occupied, so that mixing seemed to be an incomplete process. At the same time some indications of partial preservation of initial conditions in energy space were observed in N-body dissipationless simulations (for a brief review on VR as well as the issue of surviving memory of initial conditions, see \citealt[and references therein]{Dan06MNRAS}).

Indeed, it has been shown that, in general, there is a ``coarse-grained'' sense in which, after VR, the {\it ordering} of the mean energy of collections of particles is preserved, although the individual particle energies and their phase-space variables do vary substantially from their initial values (even though not in an exhaustive manner, as mentioned above). This {\it energy ranking preservation} (hereon, ERP) effect was studied by Kandrup et al. in the $90$'s, in the context of dissipationless simulations of galactic models \citep{Kan93}. Hence, it seems that particles which are closely spaced in energy space are able to mix in phase space, but with a restriction to the exchange of energy beyond certain bounds. This has been hypothesised by Kandrup et al. to be the result of some ``mesoscopic'' constraint, operative in the N-body gravitational problem.

The ERP effect is not well-understood and has not been thoroughly discussed in the literature after the passing of Kandrup in 2003 \citep{Mer05}. We have previously analysed the ERP effect in numerical studies and obtained a few hints on this mechanism, which we briefly list here: {\it (i)} the (hypothetical)  mesoscopic constraints seem to be partially violated in mergers (leading to ``fundamental plane''-like relations) but entirely operative in collapses (leading to homologous systems) \citep{Dan03}; {\it (ii)} the ERP seems to depends on halo mass in dark-matter only $\Lambda$-CDM cosmological simulations \citep{Jen98,Tho98} so that more massive haloes showed more rank preservation than less massive ones \citep{Dan06MNRAS}; {\it (iii)} 
the mesoscopic constraint seems to act at the level in which collections of particles behave dynamically as an effective particle with a characteristic frequency in the average potential of the halo \citep{Dan06CEL}.

In the present paper, we revisited the ERP effect by analysing the state of the art IllustrisTNG cosmological simulations\footnote{https://www.tng-project.org/}, produced from a sophisticated magnetohydrodynamical numerical code \citep[{\tt AREPO}, ][]{Spr10}, incorporating semi-analytical models which follow the coupled dynamics of baryons and dark matter \citep{Nel19,Pil18,Spr18,Nel18,Nai18,Mar18,TNG50-1,TNG50-2}. Our purpose was to advance the quantitative characterisation of ERP, prompted by open questions raised in \cite{Kan93} and by previous hints on this effect (mentioned above). For instance, does ERP (or its subliminal mesoscopic constraints) have some ``universal'' characterisation? In order to map this problem, a preliminary focus could be in terms of parameters involving {\it the collective range in which of ERP is effective}, such as the coarse-graining level in the energy space. Another distinct set of quantities to be correlated with ERP would be {\it global and historical characterisations} of gravitational systems evolving towards equilibrium. Cosmological haloes in the present simulations offer interesting testing grounds for such questions, not only at the fundamental level but also as potential developmental tools, when well-understood, for observational surveys.  
 
Our paper is organised as follows: in Sec. \ref{METH}, we present the selection of the IllustrisTNG runs, the methodology for computing the coarse-grained energy ranking statistics, and a description of toy models, where we illustrate the general behaviour of ERP fractions. In Sec. \ref{RES}, we present the energy ranking results and the dependence of ERP fractions with halo properties. Our results are discussed in Sec. \ref{DIS}. The Appendix provides a compilation of the ERP fractions for haloes other than the $4$ most massive ones, in each simulation (which were presented in detail in the main text). The cosmology parameters used in the present paper are those defined in the IllustrisTNG simulations, namely: $\Omega_{\Lambda,0}= 0.6911, \Omega_{{\rm m},0} = \Omega_{{\rm DM},0} + \Omega_{{\rm b},0} = 0.3089, \Omega_{{\rm b},0} = 0.0486, \sigma_{8} = 0.8159, n_s = 0.9667$ and $h = 0.6774$ (Planck Collaboration, \citealt{PLANCK16}). 

\section{Methodology}
\label{METH}

\subsection{Energy ranking algorithm and selection of simulations}
\label{ALG}

\begin{figure}
\centering
\includegraphics[trim= 0in 0in 0in 0in,clip,width=0.5\linewidth]{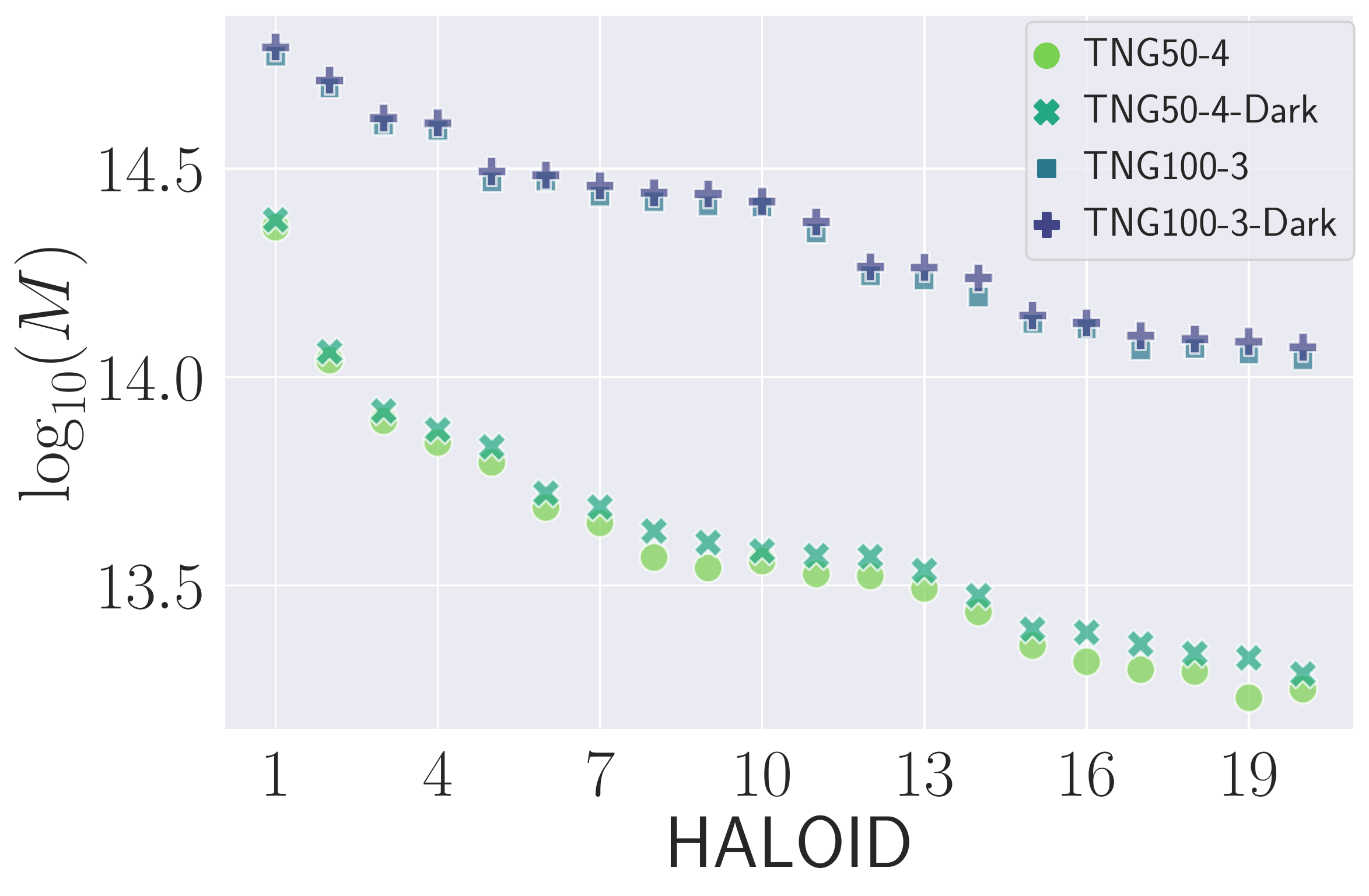} 
\caption{FOF halo masses (in $M_{\odot}$) as a function of {\tt Halo ID} for the selected simulations (namely, the $20$ most massive haloes at $z=0$  in each simulation). Masses are defined as the sum of the individual masses of all particles (of all types) associated with the FOF halo. \label{HALO-MASSES}}
\end{figure}

In this section, we describe in detail the energy ranking computation performed in the selected IllustrisTNG simulations (indicated below). A note of caution concerns our selection of haloes, to be discussed in the item (iii) below, in which we selected the $20$ most massive haloes in each simulation. Given that the TNG50 and TNG100 simulations differ in box size, the latter simulation samples larger halo masses than the former. This is seen in Fig. \ref{HALO-MASSES}. The algorithm can be stated through the following steps:

\begin{enumerate}
\item{Choose the {\it simulation} from the set: {\tt TNG} = $\{$ TNG50-4 (full), TNG50-4-Dark, TNG100-3 (full), TNG100-3-Dark $\}$. }
\item{Choose the {\it particle type} from the set:  {\tt Particle Type} = $\{$ ``dm'', ``star'' $\}$ for the full runs; for dark-only runs, {\tt Particle Type} is automatically ``dm''.  }
\item{Choose the {\it reference FOF halo ID} at $z=0$ from the set: {\tt Halo ID} = $\{ 1, 2, 3, \dots 20 \}$ (namely, the $20$ most massive haloes; c.f.  Fig. \ref{HALO-MASSES}). Note: haloes were selected from the available FOF halo catalogue at the IllustrisTNG site, which in turn were obtained from a standard friends-of-friends algorithm (linking length $b=0.2$). The latter algorithm was run on the dark matter particles (the other particle types are linked to the same groups from their nearest dark matter particle). All particles have a unique identity number throughout a given simulation run.  } 
\item{Choose the {\it number of bins} (i.e., the number of energy partitions or cells) from the set: {\tt nbins} =  $\{ 5, 10, 15,\\20, 25, 30\}$.}
\item{{\it Convention}: the energy bin labelled ``1'' refers to the most negative (most bounded) energy bin, and so on, monotonically up to the last energy bin, which refers to the less negative (less bounded) energy bin (e. g., the energy bin labelled ``15'' for {\tt nbins} = $15$).}
\item{Choose the {\it target redshift} from the set: {\tt z} = $\{ 1.0, 5.0, 10.0 \}$, also referred as ``markers''. Note: we did not trace ERP in a fine grid in redshift due to two reasons. First, given that the discriminating feature is the occurrence itself of rank violation across a long period in the evolutionary history of a halo, it was sufficient to detect ``crossing-over'' events of energy bins at a few ``check points'' (redshifts). Granted, energy ranking might proceed differently in between any such markers (i.e., with bin preservations and violations changing during that time), but the point was the detection of violations {\it per se} (relatively to the energy bin ordering at $z=0$). That is, it was not essential at this level of the analysis to trace ERP along small intervals in {\tt z}. A detailed study of the evolution of energy cells, in relation to specific dynamical/structural halo parameters, would certainly benefit from a finer partition in redshift and with the analysis of higher resolution simulations (such as the TNG50-1). However, the present investigation was focused on a more generic characterisation of ERP.  Second, a high computational cost in the particle matching algorithm was found when considering several snapshots and a large number of haloes. This issue was left for a future work, with expected improvements in the algorithm and on computational availability.}
\item{For a given choice [{\tt TNG}, {\tt Particle Type}, {\tt Halo ID}], obtain the  unique {\it identification number} of each particle belonging to {\tt Halo ID} at $z=0$: the set $\mathbb{I}_{\rm Halo~ID}(z=0) = \{ $ {\tt Particle IDs} $\} \in \mathbb{N}$. Note that the {\tt Particle IDs} are fixed in all redshifts and are provided by the IllustrisTNG snapshot data.}
\item{For a given choice [{\tt TNG}, {\tt Particle Type}, {\tt Halo ID}, {\tt z}], and for all FOF haloes at {\tt z}, search for the same {\tt Particle IDs} that were stored in the set $\mathbb{I}_{\rm Halo~ID}(z=0)$.  {\it Note:} not all particles in the set $\mathbb{I}_{\rm Halo~ID}(z=0)$  will be matched in FOF haloes at {\tt z}, because some of those particles will be in the field, and are here disregarded (see note below on this criterion). The number of matched particles, therefore, will generally vary at each {\tt z}. This gives the sets of matched particles at {\tt z}: $\mathbb{I}_{\rm mat}(${\tt z}$) \subseteq \mathbb{I}_{\rm Halo~ID}(z=0)$.}
\item{For each set $\mathbb{I}_{\rm mat}(${\tt z}$)$, store the corresponding one-particle energies at $z=0$ and at {\tt z}. The one-particle energy is calculated as the sum of the kinetic energy (in units of $(km/s)^2a$, where $a$ is the scale factor at the given redshift) and the gravitational potential energy (in units of $(km/s)^2/a$) of the particle, expressed as physical quantities by considering the respective $a$ values, provided by the IllustrisTNG specifications.}
\item{For each set $\mathbb{I}_{\rm mat}(${\tt z}$)$: {\it (a)} sort the one-particle energies at $z=0$ only; {\it (b)} assign an equal number of particles into {\tt nbins} energy cells, resulting in a sequence of energy bins, from the most bounded to the less bounded energy bin; and {\it (c)} calculate the average one-particle energies per bin, $\langle E (\rm bin) \rangle (z=0)$.}
\item{For each set $\mathbb{I}_{\rm mat}(${\tt z}$)$: assign the matched particles to the same energy bins as of the previous item and calculate their respective average one-particle energies per bin. However, these new energy bins will generally result in different average particle energies per bin, namely, $\langle E ({\rm bin}) \rangle (${\tt z}$) \neq \langle E ({\rm bin}) \rangle (z=0)$. Most importantly, there is no a priori reason that these new energy bins will follow the same order as those at $z=0$ but, if so, we have ERP.}
\end{enumerate}

We point out that our procedure follows essentially the same algorithm of \cite{Kan93}, with the difference that we considered particles {\it belonging to FOF haloes only}, disregarding field particles. Note that a reference halo at $z=0$ usually will have several particles which are not found in haloes at higher redshifts, and therefore these particles were not considered. The reason for this decision was due to computer memory limitations, as loading the whole snapshot of the simulation box would be required. Some solutions to this problem implied a significantly more complex algorithm, which we left for future work. On the other hand, by focusing on particles associated with FOF haloes only, the resulting analysis corresponds more closely to what is actually found in observational surveys, as the latter ultimately sample bound structures, given the inevitable flux limitations. Therefore, by limiting to halo particle data only, we were able to bypasses computer memory limitations, while providing a starting point for subsequent, more refined observational comparisons with simulated data.

\subsection{ERP diagrams}
\label{DIAG}

\begin{figure}
\centering
\includegraphics[width=0.7\linewidth]{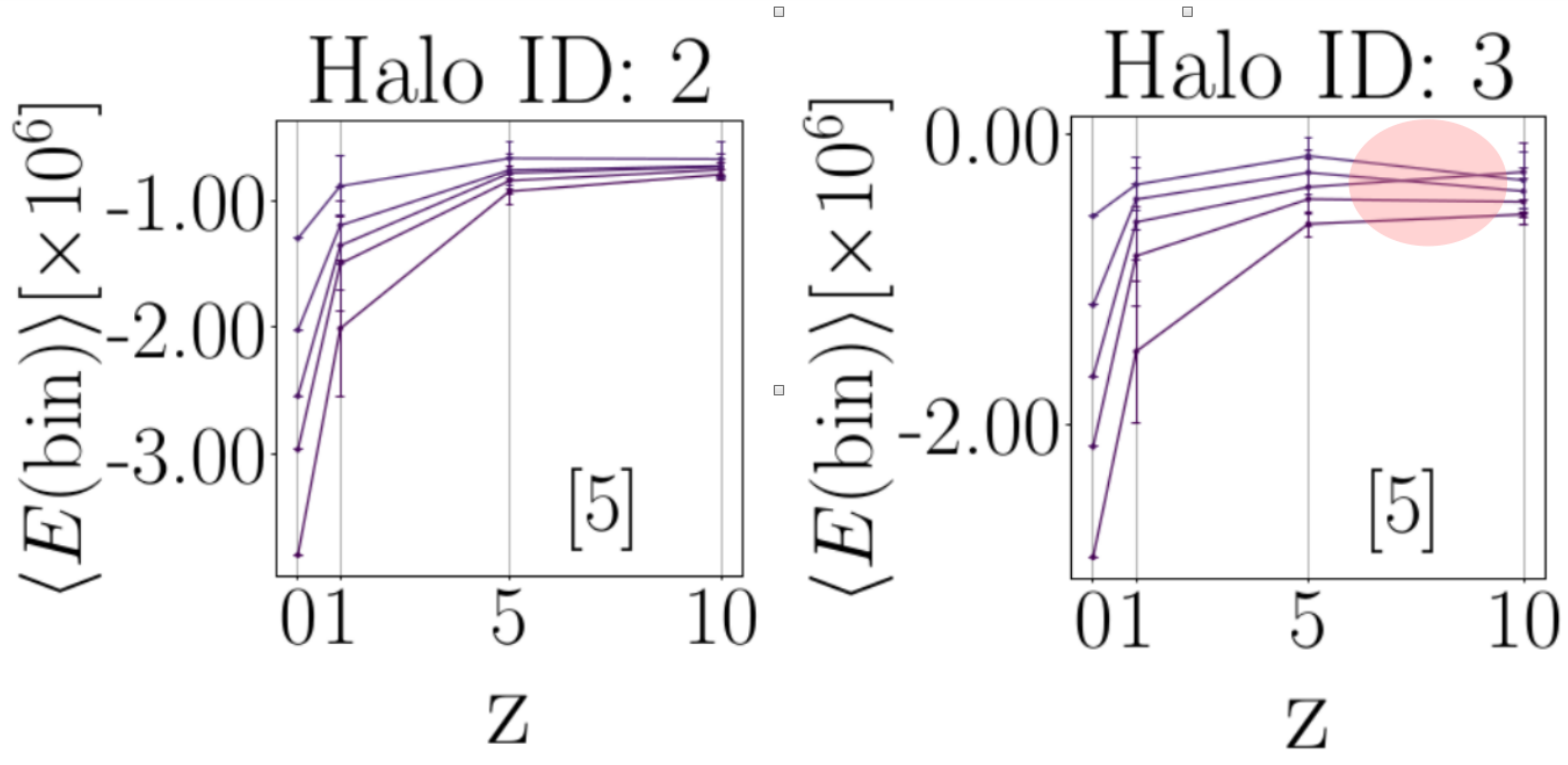}
\caption{Example: ERP ``criss-cross'' diagrams, for {\tt nbins} = $5$. Mean energies $\langle E({\rm bin}) \rangle$ were computed for reference {\tt Halo ID} $=2$ (left panel), and  reference {\tt Halo ID} $=3$ (right panel), both for TNG100-3, {\tt particle type} = ``dm''. The variances of particle energy values relative to each $\langle E({\rm bin}) \rangle$ are shown as error bars at the target redshifts {\tt z} $= \{ 1.0, 5.0, 10.0 \}$. Note the complete ERP in the left panel and $3$ bin violations in the right panel (indicated by an ellipse), at {\tt z} $=10.0$; the corresponding rank-violating bin levels were $\ell_{\rm viol} = 3, 4, 5$. Higher bin levels correspond to less bounded (less negative) energy bins. \label{ERPExamples}}
\end{figure}

\begin{figure}
\centering
\includegraphics[trim= 0in 0in 2.9in 0in,clip,width=0.7\linewidth]{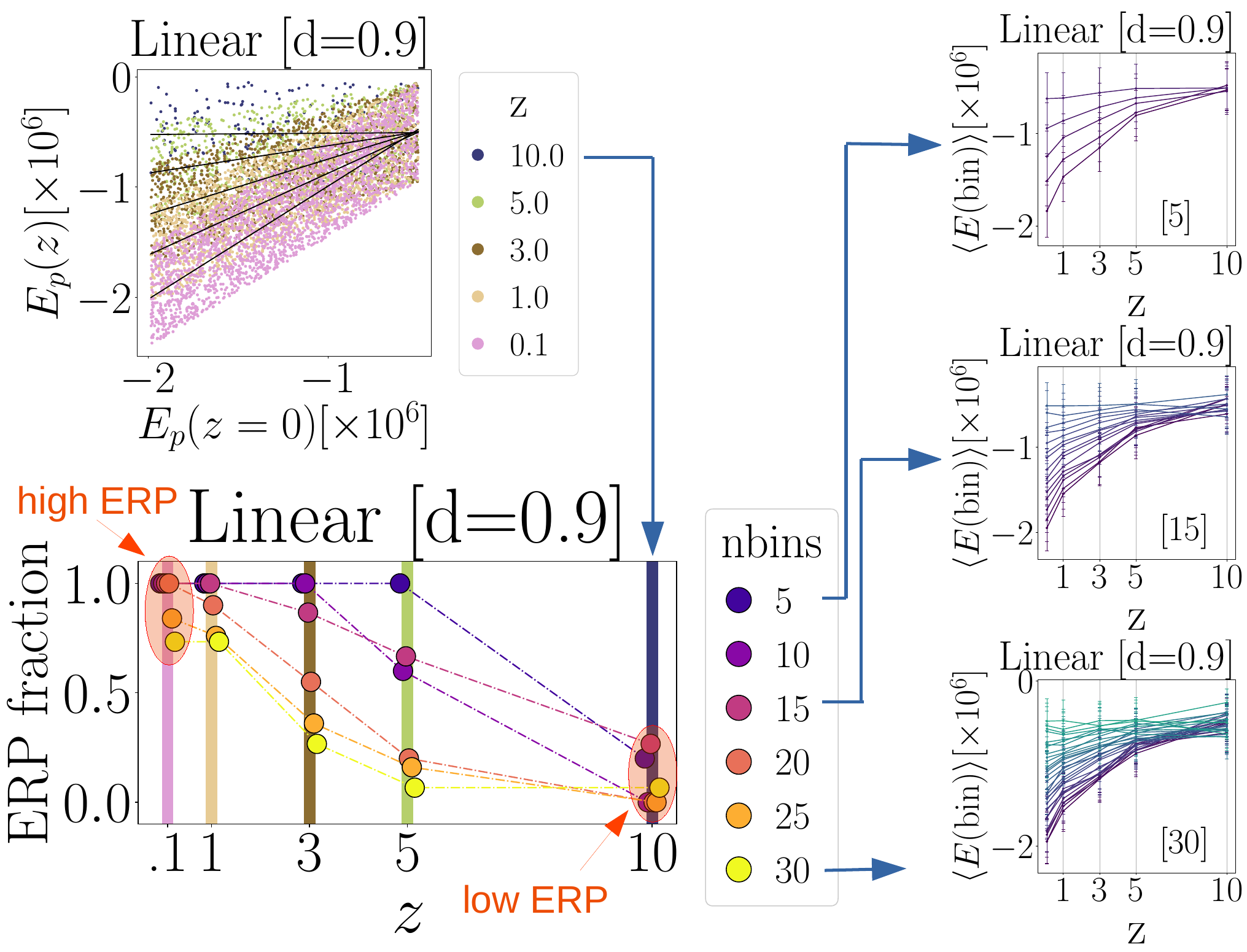} 
\caption{Example: ERP analysis of a linear toy model, illustrating the particle correlation space $E(z=0) \times E(z)$ (top left panel); the ``criss-cross'' diagrams, in which only $3$ values of {\tt nbins} are shown inside brackets, namely: {\tt nbins} $= \{5, 15,30 \}$ (right panels; {\tt z} $=0.1$ axis label omitted for clarity); and the corresponding ERP fractions (with data points slightly jittered for a better view of overlapping points; bottom left panel). The number of violated bins  ($n_{\rm viol}$) for this model were the following. 
For {\tt nbins} $=5$: $n_{\rm viol} = 4$ (at {\tt z} $=10.0$);
for {\tt nbins} $=15$: $n_{\rm viol} = 11,5,2$ (at {\tt z} $=10.0,5.0,3.0$, respectively);
for {\tt nbins} $=30$: $n_{\rm viol} = 28,28,22,8,8$ (at {\tt z} $=10.0,5.0,3.0,1.0,0.1$, respectively). \label{ILL-EPRF}}
\end{figure}

The algorithm described in the previous section indicates whether the ranking of {\it the centres of the energy cells} are preserved (equivalently, the ordering of the computed $\langle E (\rm bin) \rangle (${\tt z}$)$), relatively to their values at $z=0$. Energy ranking results can be compactly presented in a ``criss-cross''  diagram, which can be promptly visualised. This is done by connecting the $\langle E({\rm bin}) \rangle(z)$ results by line segments, one for each average energy bin. We show an example of our results (to be discussed in Sec. \ref{RES}) in Fig. \ref{ERPExamples}: pictorially, this means that there are no traversing of any line segments across redshifts, {\it (a)}, whereas some energy ranking violations are seen in {\it (b)}, where crossings over line segments were found.

However, ``criss-cross''  diagrams will be presented here in a limited manner and mostly for illustrative purposes, considering its connection to previous literature. The large number of ERP evaluations generated $720$ ``criss-cross''  diagrams, due to our combinatorial choices in {\tt [TNG, Particle Type, Halo ID, nbins]} (c.f. previous subsection). It was necessary to reduce this large ERP information, so we calculated the {\it ERP fractions} (ERPFs) at each redshift. That is, the fraction of bins that were {\it preserved}, relatively to the evaluated {\tt nbins}, at {\tt z}. Hence, ERPF = $1.0$ indicates complete energy ranking preservation of all bins at a given {\tt z}, whereas ERPF = $0.0$  means a complete violation of all bins at a given {\tt z}. ERPF can therefore be compactly displayed for all {\tt nbins} at {\tt z}. This is illustrated in Fig. \ref{ILL-EPRF}, obtained for a toy model (to be discussed next). One can see then that ERP information for all {\tt nbins} values can be compactly displayed in one diagram.

Granted, ERPF leads to loss of information on the specific bins that are violated. Yet, we believe that the ERPF can be considered a sufficiently meaningful (albeit reduced) quantity, and it can be directly correlated with other halo properties (c.f. Sec. \ref{RES-ERPF}). Another caveat is that the presence of overlapping variances in the ERP ``criss-cross''  diagrams could be considered as statistical measures of potential bin violations. Given that the main driver of ERP has been observed in the energy bin averages, and not in the variances \citep{Kan93}, we assume hereon that our ERPF analysis is meaningful in a first-order sense.

\subsection{Toy Models and general behaviour of ERP fractions}
\label{TOY}

\begin{center}
\begin{table}
\centering

   \begin{tabular}{ | l | l | c | c | c } \hline \hline
    Energy correlation model  & {\tt d} factor & $a$      & $b$ & $\mathfrak{z} $  \\ \hline \hline
    Linear:  $\mathfrak{E}(\mathfrak{z})=aE_0+b$  & $\{0.1,0.3,0.5,0.7,0.9 \}$ &  &  & \\
             & & $0.00$  & $-5.00\times 10^5$ & $10.0$\\
             & & $0.25$  & $-3.75\times 10^5$ & $5.0$\\
             & & $0.50$  & $-2.50\times 10^5$ & $3.0$\\
             & & $0.75$  & $-1.25\times 10^5$ & $1.0$\\
             & & $1.00$  & $-4.30$ & $0.1$\\ \hline
    Quadratic:  $\mathfrak{E}(\mathfrak{z})= (1/b)E^2_0+ aE_0 + b$  & $\{0.5,1.0,2.0,5.0,10.0 \}$ &  &  & \\
             & & $-5.00$  & $-5.00\times 10^5$ & $10.0$\\
             & & $-4.00$  & $-5.00\times 10^5$ & $5.0$\\
             & & $-3.00$  & $-5.00\times 10^5$ & $3.0$\\
             & & $-2.00$  & $-5.00\times 10^5$ & $1.0$\\                                      
             & & $-1.00$  & $-5.00\times 10^5$ & $0.1$\\ \hline
 \end{tabular} 
\caption{ Energy correlation models (linear and quadratic), resulting in two sets of $5$ toy models each, produced according to the listed values of the disturbance factor ({\tt d}). At each artificial redshift $\mathfrak{z}$, the adopted parameter values $a$ and $b$ are indicated, aimed to produce increasingly correlated particle energies with ``time''. }
\label{TAB-MODELS}
\end{table}
\end{center}

Before we proceed to the analysis of the simulation results, we introduce two sets of toy models to serve as a basis for the understanding of how energy ranking is best preserved or otherwise gradually violated in terms of general trends. These models are completely artificial in the sense that we generated dummy particle energies at arbitrarily labelled redshifts, that is, not based on any cosmological considerations, but constructed in a specified manner. They should be seen as useful references for the qualitative understanding of the energy ranking behaviour as a function of certain broad characteristics. The rationale for tracing the ERP in such a well controlled dataset is due to the fact that the main quantity being computed in the cosmological simulations is the {\it ordering} of energy cells, and not the specific values of particle energies. Therefore it is possible to capture the main dependencies of the degree of ERP directly from the broad characteristics in the particle correlation space $E(z=0) \times E(z)$, as we illustrate now.

We produced models in which we fixed the particle energies $E({\rm z=0}) \equiv E_0$ for the choice {\tt [TNG, Particle Type, Halo ID]} = [TNG50-4-Dark, ``dm'', 1], and then generated artificial particle energies for dummy redshifts ($\mathfrak{z}$), denoted $\mathfrak{E}(\mathfrak{z})$, based on two parameterised correlations (linear and quadratic), with a spread around each correlation, specified by a disturbance factor ({\tt d}). That is, after obtaining a particle energy $\mathfrak{E}(\mathfrak{z})$, the latter was disturbed by a value drawn from a uniform distribution within the interval ${\tt d}$, namely: 
$\mathfrak{E}(\mathfrak{z}) = \mathfrak{E}(\mathfrak{z}) \pm (-5.00\times 10^5\times${\tt d}$)$, where the sign $\pm$ was also chosen randomly. In this manner, each disturbance factor created a different realisation of the model, with a sequentially increased spread around the given correlation. Furthermore, we also mimicked the decrease of bounded particles in haloes at higher redshifts by randomly sampling the generated particle energies from arbitrarily smaller numbers at higher redshifts.   A summary of $10$ toy models generated in that manner is presented in Tab. \ref{TAB-MODELS}.

The idea is to understand the impact of the type and degree of correlation, its spread, and the number of energy bins on the corresponding ERP fraction at each redshift. The complete results for the toy models are presented in Fig. \ref{TOY-BOTH}.  The models were constructed in such a way that in the particle correlation space, particle energies are increasingly correlated (the slope increases from high to low redshifts). Given that the models are intended to capture the main trends of ERP in terms of correlation level, spread and {\tt nbins}, we list the following observed properties, which can be taken as a reference for simulated data:

\begin{figure}
\centering
\includegraphics[trim= 0in 0in 0in 0in,clip,width=1.0\linewidth]{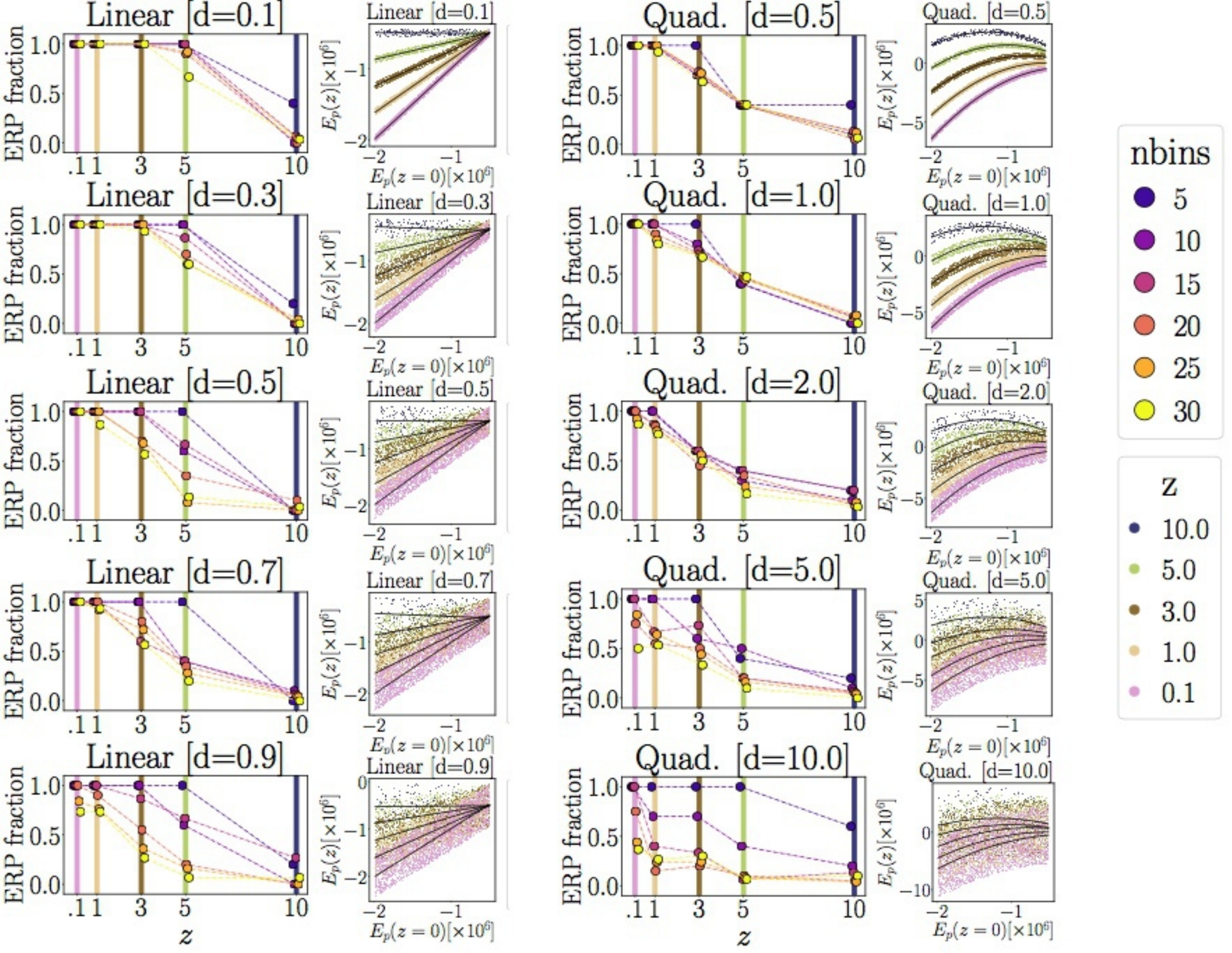}
\caption{Summary of the ERP results for the toy models described in the main text. \label{TOY-BOTH}}
\end{figure}

\begin{itemize}
\item{Correlation: As qualitatively expected, there is a strong connection between ERP fraction and the degree of correlation between the particle energies, $E(z=0)$ and $E(z)$, whatever the spread of the correlation and the {\tt nbins} used. This can be seen clearly for the case of linear models, where lower redshifts (higher slopes) lead to higher ERP fractions. The quadratic models also show the same trend, but ERP is more compromised overall, as the curves present only a certain range with an almost linear correlation. }
\item{Spread: the ERP fractions in both correlation models tend to be lower for higher spreads, but as mentioned in the previous item, this is a second order effect as compared to the degree of the correlation. Therefore the amount of spread on generating lower ERP fractions is more important for less correlated particle energies (at higher redshifts). Quadratic models show more sensitivity on spread for all redshifts than linear models.}
\item{{\tt nbins}: By focusing on the two uppermost panels of the linear model and the three uppermost ones of the quadratic model, that is, models with less spread, we see that they present a remarkable {\it consistency} of EPR fractions, regardless of the choice of {\tt nbins}. }
\end{itemize}

\section{Results}
\label{RES}
\subsection{ERP Results}
\label{ERP-RES}

\begin{figure}
\centering
\includegraphics[trim= 0in 0in 0in 0in,clip,width=0.6\linewidth]{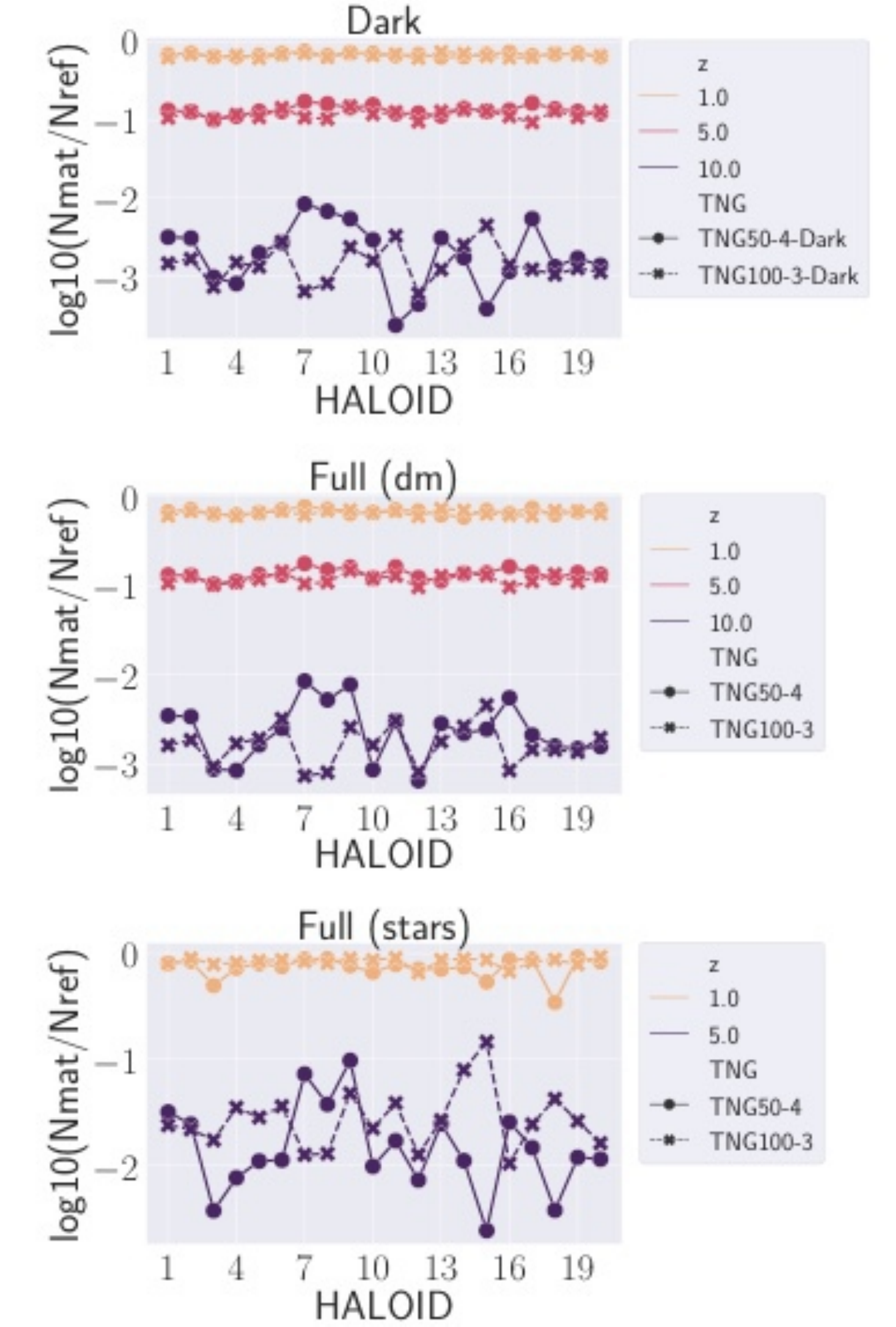}  
\caption{Fractions (in logarithmic scale) of matched particles to total particles (in the respective reference haloes), for each target redshift {\tt z} and {\tt Particle Type}, for all analysed simulations. \label{MATP}}
\end{figure}

\begin{figure}
\centering
\includegraphics[trim= 0in 0in 0in 0in,clip,width=0.76\linewidth]{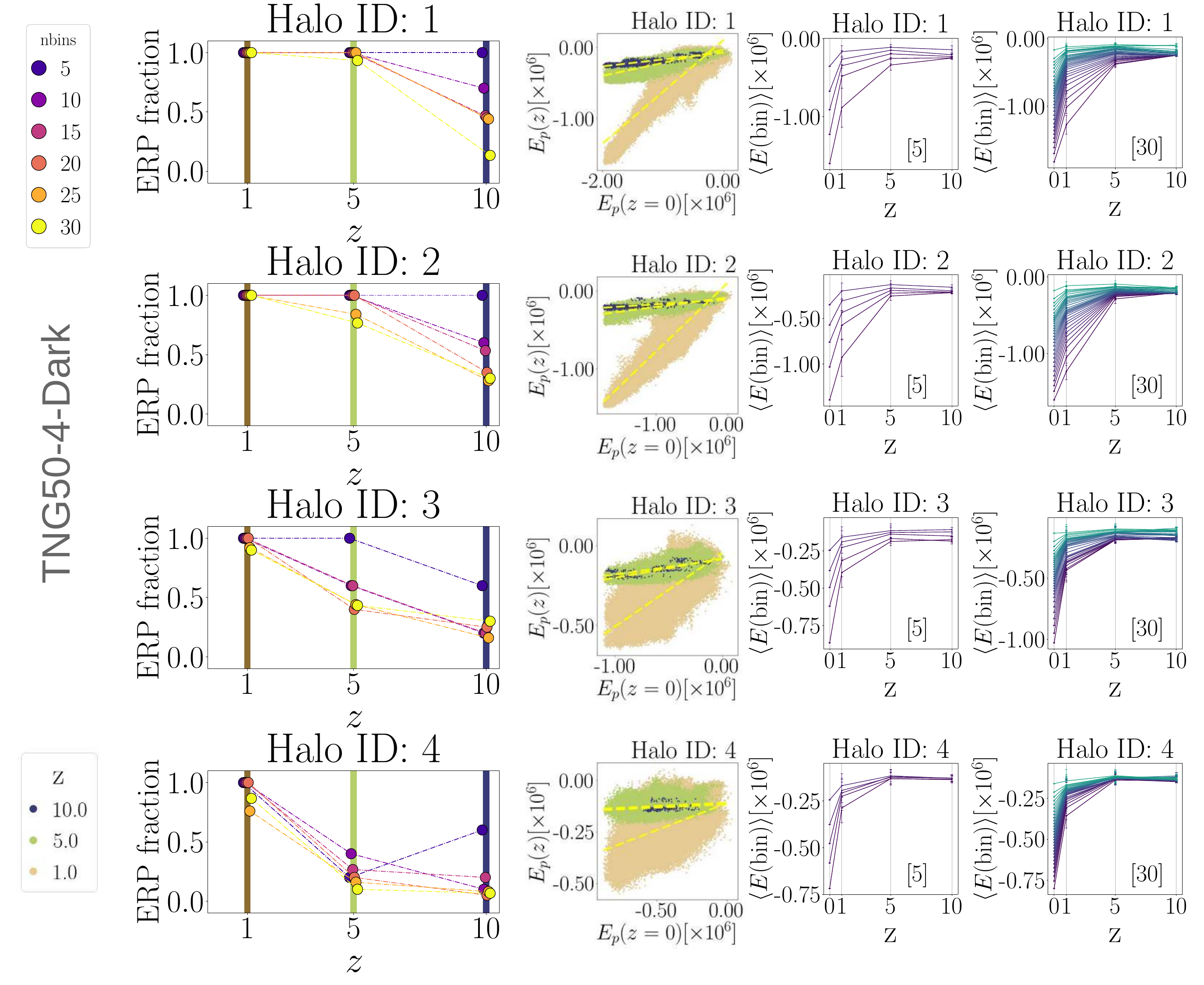}\\  
(a) \\ $~$ \\
\includegraphics[trim= 0in 0in 0in 0in,clip,width=0.76\linewidth]{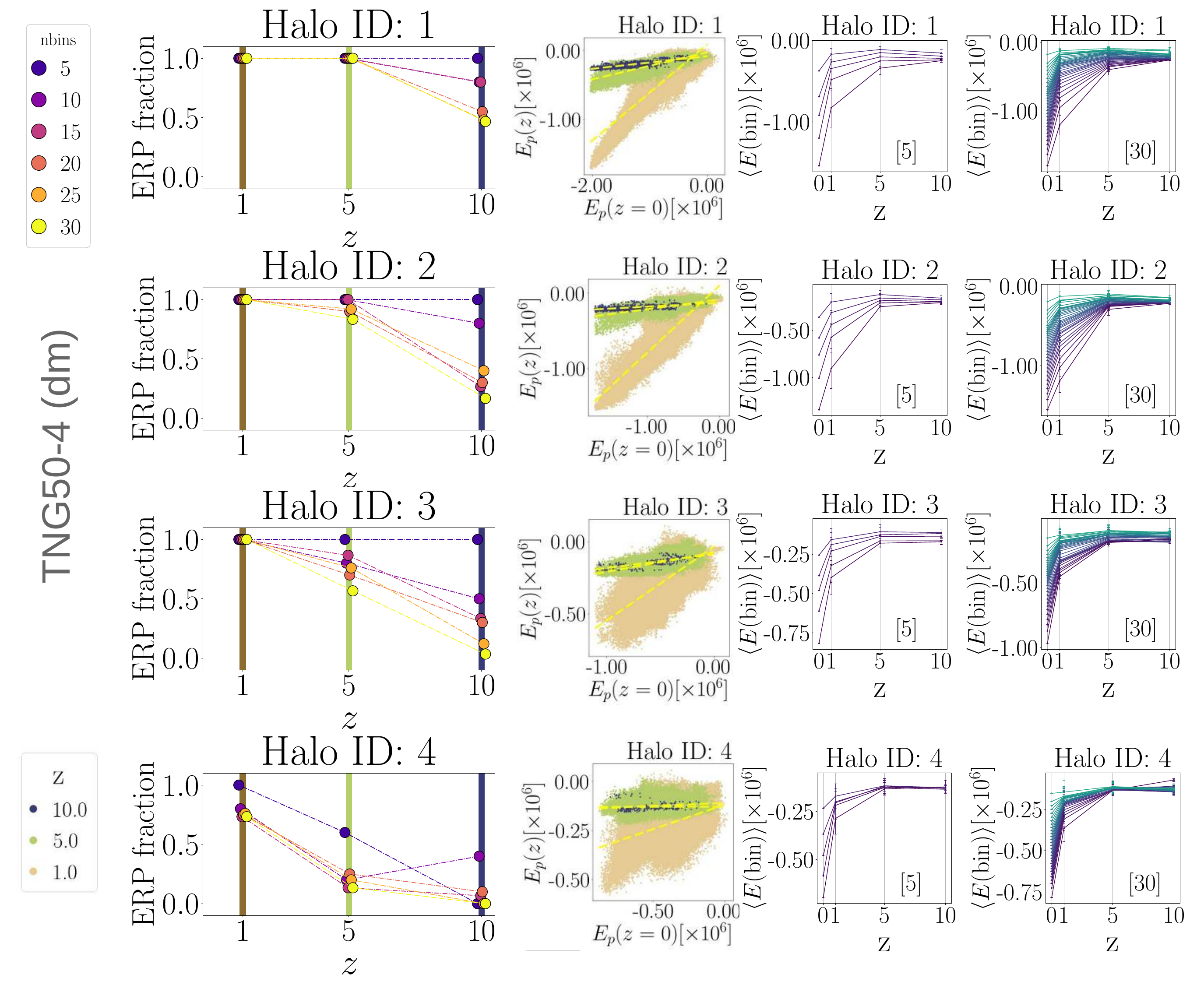}\\  (b)
\caption{ERP results for the (a) TNG50-4-Dark and (b) TNG50-4 full ({\tt Particle Type:} ``dm'') simulations, {\tt Halo ID} $= \{1, \dots 4 \}$.  {\it Left panels:} ERP fractions; {\it middle panels:} particle correlation spaces (dashed lines: linear fits); {\it right panels:} ``criss-cross'' diagrams (only {\tt nbins} $= \{ 5, 30 \}$ are shown). \label{EPR_PANELS-T50}}
\end{figure}

\begin{figure}
\centering
\includegraphics[trim= 0in 0in 0in 0in,clip,width=0.77\linewidth]{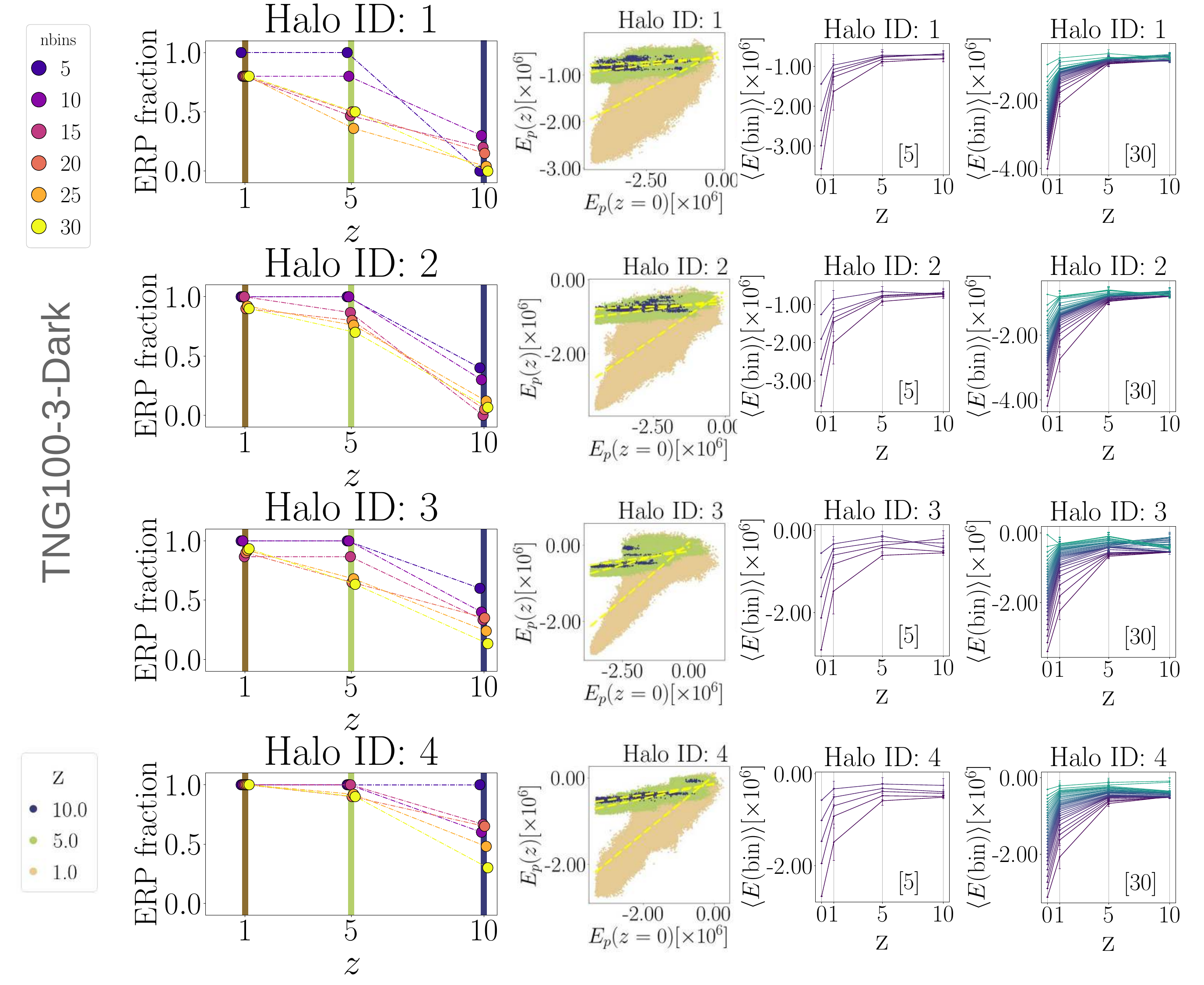}\\  
(a) \\ $~$ \\
\includegraphics[trim= 0in 0in 0in 0in,clip,width=0.77\linewidth]{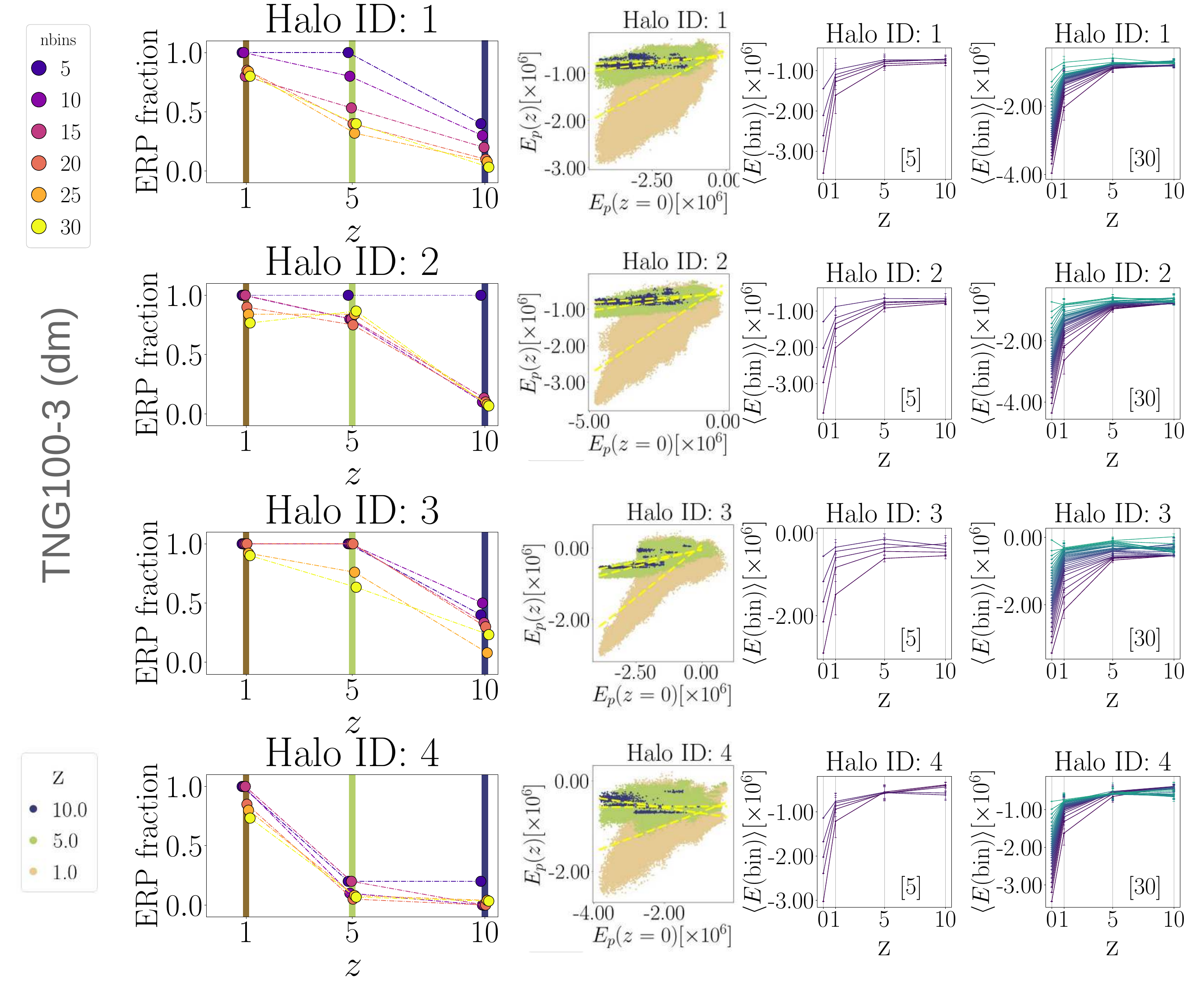}\\ (b)
\caption{Same as previous figure, for the (a) TNG100-3-Dark and (b) TNG100-3 full ({\tt Particle Type:} ``dm'') simulations. \label{EPR_PANELS-T100}}
\end{figure}

\begin{figure}
\centering
\includegraphics[trim= 0in 0in 0in 0in,clip,width=0.77\linewidth]{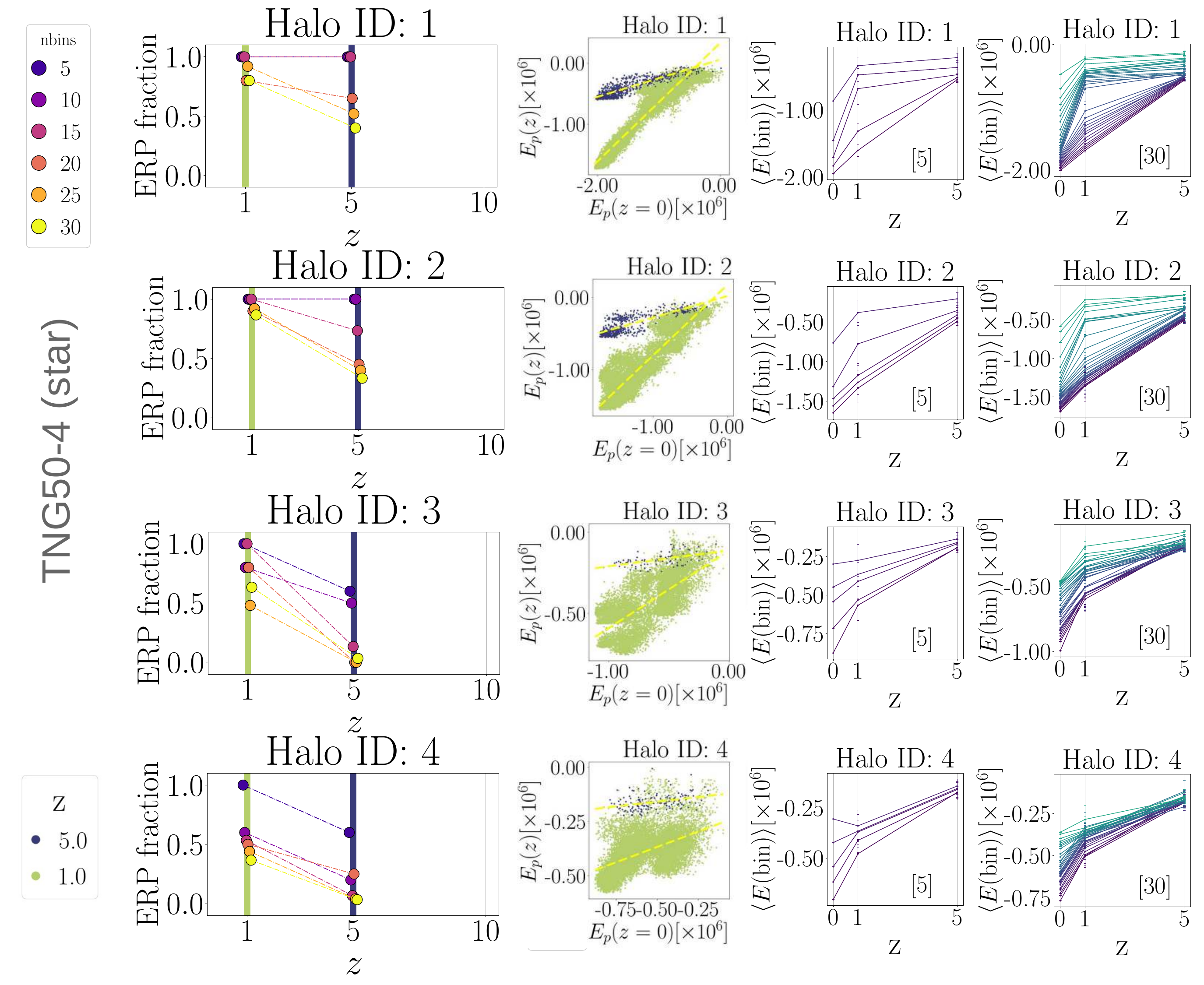}\\  
(a) \\ $~$ \\
\includegraphics[trim= 0in 0in 0in 0in,clip,width=0.77\linewidth]{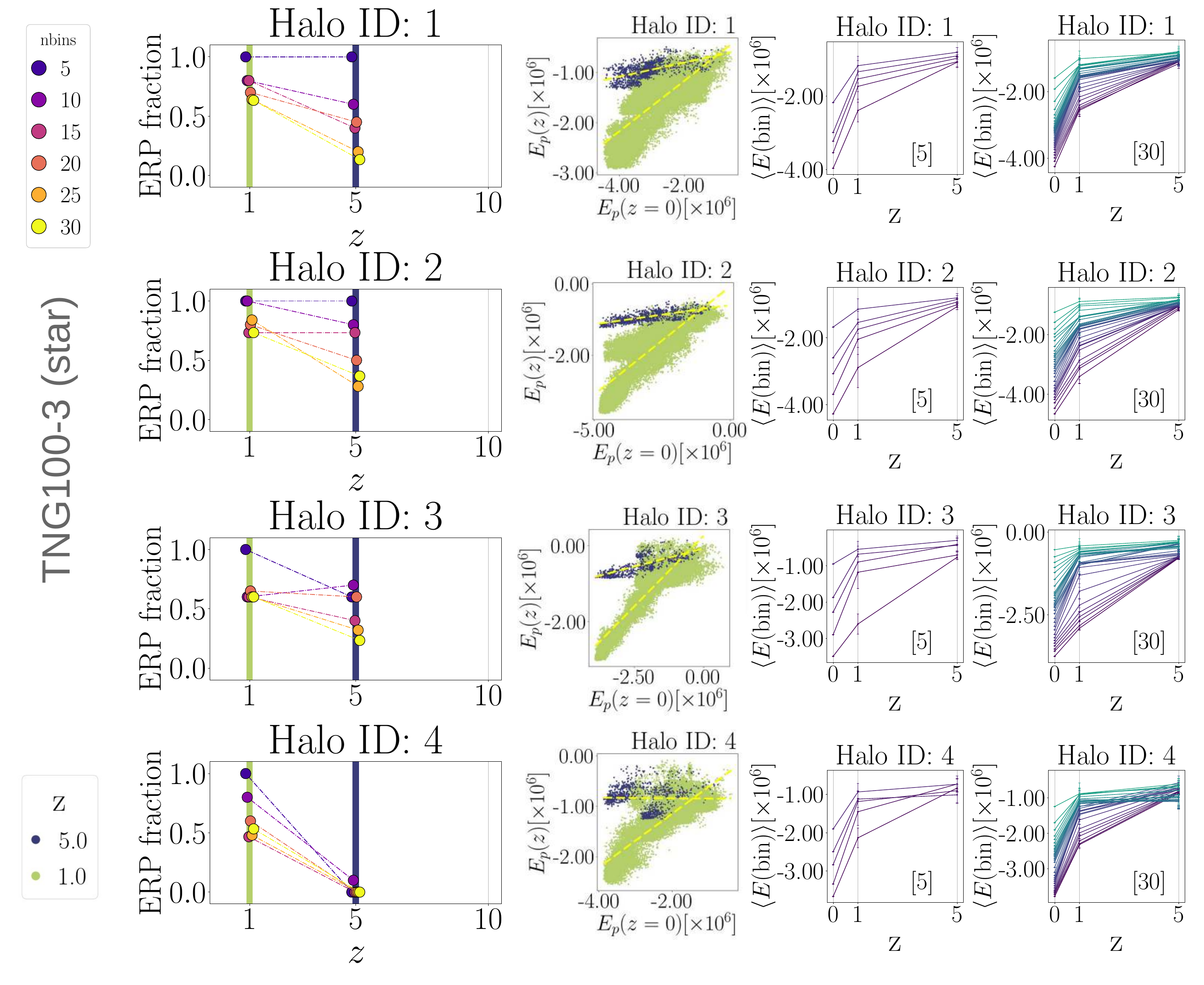}\\ (b)
\caption{Same as previous figure, for the (a) TNG50-4 full and (b) TNG100-3 full (both, {\tt Particle Type:} ``star''). \label{EPR_PANELS-st}}
\end{figure}

\begin{figure}
\centering
\includegraphics[trim= 0in 0in 0in 0in,clip,width=0.6\linewidth]{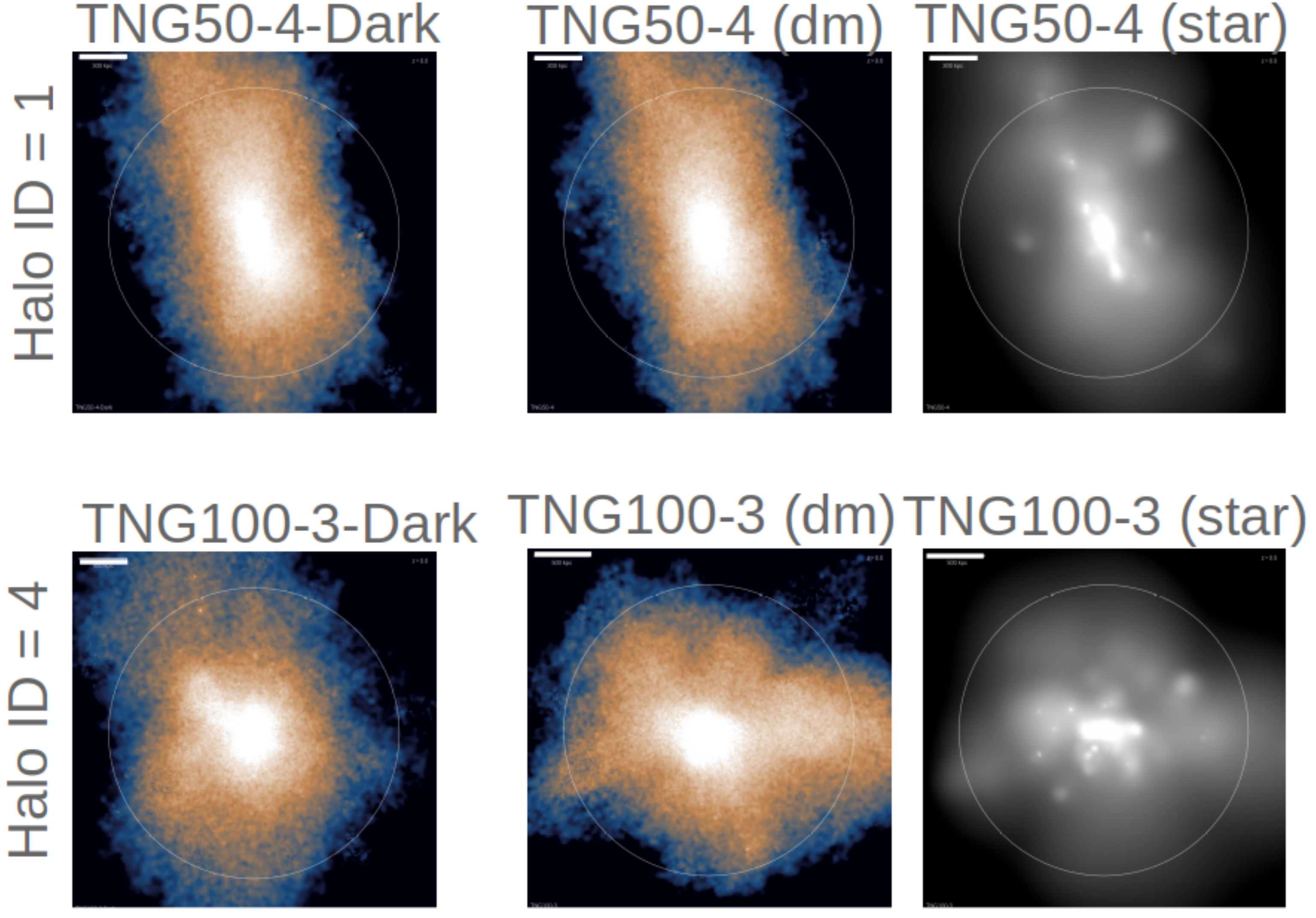}
\caption{2D projections of the dark matter column densities ($\log_{10} M_{\odot} {\rm kpc}^{-2}$) for the two examples in which there ERPF results are similar (TNG50-4, {\tt Halo ID} $= 1$) and different (TNG100-3, {\tt Halo ID} $= 4$). Stellar column densities ($\log_{10} M_{\odot} {\rm kpc}^{-2}$) are also presented at the respective right panels. Column density ranges are (from left to right): TNG50-4 = $[5.8, 8.4]$,$[5.8, 8.4]$, $[2.0,6.0]$; TNG100-3 = $[6.3, 8.5]$,$[6.0, 8.4]$, $[2.2,6.0]$. These figures were made using the web-based API functionality; circles indicate the respective virial radii. Scale bars are: $300$ kpc for TNG50-4 (three upper panels); $400$ kpc for TNG100-3-Dark and $500$ kpc for the remaining two TNG100-3 panels.  \label{VIS}}
\end{figure}

We show in Fig. \ref{MATP} the fractions of matched particles to total particles (in the respective reference haloes). Those fractions are systematically lower for higher redshifts, as expected, and are roughly similar when comparing the TNG50 and the TNG100 simulations.
In Figs. \ref{EPR_PANELS-T50}, \ref{EPR_PANELS-T100} and \ref{EPR_PANELS-st} we present a summary of the ERP results for the first $4$ more massive reference haloes, in each of the analysed simulations. From these figures, one can inspect and compare the ERP fractions (ERPF), particle correlation spaces ($E(z=0) \times E(z)$) and ``criss-cross'' diagrams (for {\tt nbins} $= \{ 5, 30 \}$ only).  Due to space constraints, we included in the Appendix the ERPF results for all the remaining reference haloes. Note that the panels showing the particle correlation spaces ($E(z=0) \times E(z)$) indicate a decrease of the number of matched particles with {\tt z}, given the adopted criterium of choosing only particles belonging to FOF haloes, disregarding field particles.  For the case of {\tt Particle Type} = ``star'', a poor statistics was found for very high redshifts ({\tt z} $=10.0$), so we limited the analysis for the reference redshifts {\tt z} $= \{ 1.0 ,5.0 \}$ in this case. 

We found that all the ERP results were consistent with the trends described for toy models in Sec. \ref{TOY}. Clearly, one-particle energies did become increasingly correlated as one approached smaller redshifts (an effect included by construction in the case of the toy models). Interestingly, the correlations were better fit by a linear relation. However, a significant spread around the correlations were found in many cases, so other relations could have been fitted as well. This spread indicated therefore a partial loss of ``memory'' across the evaluated redshifts. Yet, such a mixing (in the one-particle energy space) was not necessarily as complete as one would infer: even those systems with high scatter showed significant ERPFs, specially at {\tt z} $= 1.0$ (i.e., ERPF greater than $\sim 0.8$ for {\tt nbins} at least as large as $15$). For instance, see {\tt Halo ID}s $= \{ 3, 4 \}$ for the TNG50-4 (``dm'') in Fig. \ref{EPR_PANELS-T50}, and {\tt Halo ID}s $= \{ 2, 3 \}$ for the TNG100-3 (``dm'') in Fig. \ref{EPR_PANELS-T100}.

The results for {\tt Particle Type:} ``star'' (Fig. \ref{EPR_PANELS-st}) were less robust due to poorer statistics at higher redshifts, yet, we found indications of ERP for this component as well, although in stricter conditions (overall, for {\tt z} $=1.0$ and {\tt nbins} $=5$). There are some cases in which ERPF is quite high (greater than $\sim 0.8$) even for {\tt nbins} as high as $=30$ (fixing {\tt z} $=1.0$); see, e.g., {\tt Halo ID}s $= \{ 5, 6, 8, 9, 13, 14, 16, 20 \}$ (TNG50-4); and {\tt Halo ID}s $= \{ 5, 8, 10, 11, 12, 15, 19 \}$ (TNG100-3) in the Appendix. Some cases of high EPRFs at {\tt z} $=5.0$ and {\tt nbins} $=5$ can also be found by inspection of these diagrams.

A direct comparison of Dark vs. full (``dm'') halo pairs showed that the EPRFs were qualitatively similar, indicating lack of sensitivity of dissipative processes. However, some exceptions were noted, specially at {\tt z} $=5.0$ and/or at higher {\tt nbins}. Namely, somewhat differing results in the energy bin levels in those pairs were seen in higher redshifts for: TNG50-4,  {\tt Halo ID}s $= \{ 3, 9, 10,11,15,16, 17, 18, 19, 20\}$; and TNG100-3, {\tt Halo ID}s $= \{ 4, 5, 6, 7, 8, 9, 11, 13, 14, 16, 18, 19\}$ (c.f. Figs. \ref{EPR_PANELS-T50}, \ref{EPR_PANELS-T100} and the Appendix). Despite such deviations, they showed an overall similarity, specially for lower redshifts and/or smaller {\tt nbins}. An extreme example was TNG100-3, {\tt Halo ID} $= 4$: the scatter in the $E(z=0) \times E(z)$ space was clearly larger in the full run, in comparison to its Dark counterpart (c.f. Fig. \ref{EPR_PANELS-T100}). For illustrative purposes, we show in Fig. \ref{VIS} visual 2D projections of the dark matter column densities for the two examples in which there ERPF results are similar (TNG50-4, {\tt Halo ID} $= 1$) and the latter extreme case (TNG100-3, {\tt Halo ID} $= 4$). Stellar column densities are also presented at the respective right panels in this figure. It is interesting to note that, for the TNG50-4 case, the dark matter column densities were quite similar, from visual inspection. However, the situation for the TNG100-3 showed a significant difference between both dark matter column densities (Dark vs. full run), with the stellar counterpart following the dark matter distribution of the full run. It is also interesting to note that the results for the {\tt Particle Type:} ``star'' in the TNG50-4 case ({\tt Halo ID} $= 1$) showed higher ERPFs than for the TNG100-3 ({\tt Halo ID} $= 4$).

In general, the simulation results were consistent with the trends found in the toy models: the ERP fractions tended to be lower for higher spreads, but this appears to be a second order effect as compared to the degree of the correlation.  In other words, the particles partially mixed in the one-particle energy space, with their collective energy rankings remarkably preserved at lower redshifts. The highest redshift {\tt z} $=10$ resulted, overall, in the lowest ERPFs, as expected. Yet, for {\tt nbins} $=5$ (or even more finely grained partitions), one still could find cases of complete ($100 \% $) ERPFs at those high redshifts, which is worth noting.

\subsection{Dependence of ERP fractions with halo properties}
\label{RES-ERPF}

\begin{figure}
\centering
\includegraphics[trim= 0in 0in 0in 0in,clip,width=0.6\linewidth]{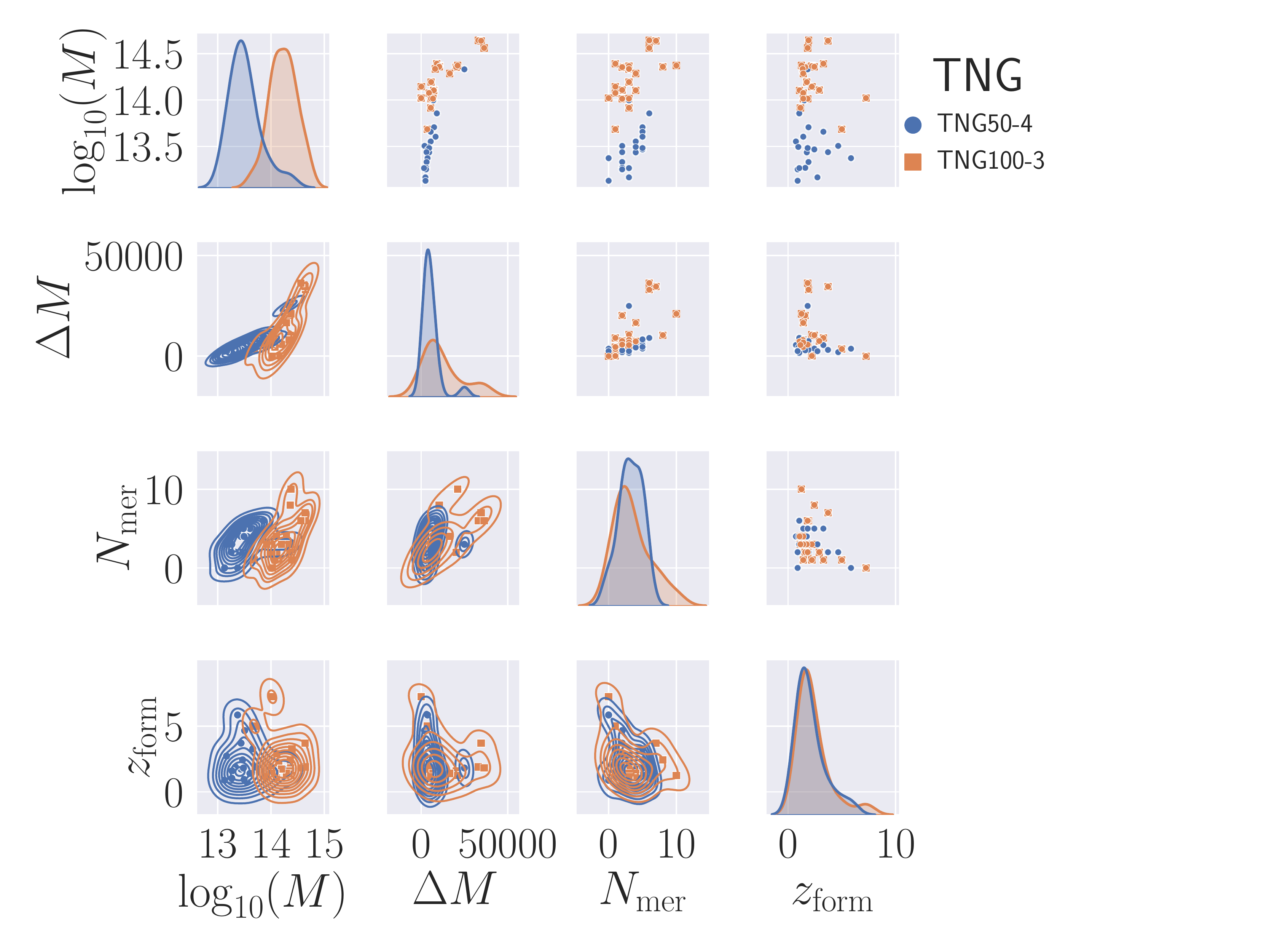}
\caption{Paired density and scatterplot matrix for TNG50-4 and TNG100-3, for the following variables: total reference halo mass at $z=0$ (in $M_{\odot}$; logarithmic scale); fractional increase in halo mass, $\Delta M$; number of major mergers, $N_{\rm mer}$, and formation redshift, $z_{\rm form}$. \label{Pairplot_haloprops}}
\end{figure}

In this section we present a study of the behaviour of the ERP fractions as a function of certain halo properties. Our purpose was to advance the ERP methodology as a tool for the theoretical study of mixing in haloes, by connecting it with halo global properties and historical markers. For the present analysis, we considered: {\it (i)} total reference halo mass at $z=0$; {\it (ii)} fractional increase in halo mass, $\Delta M$;  {\it (iii)} number of major mergers, $N_{\rm mer}$; and {\it (iv)} formation redshift, $z_{\rm form}$.  We defined the fractional increase in halo mass as: $\Delta M \equiv (M_{\rm f} - M_{\rm i})/ M_{\rm i}$, where the subscripts ``i'' and ``f'' refer to the initially identified progenitor and the final reference halo at $z=0$, respectively. We identified the number of major mergers along the tree for each reference halo, defining a ``major merger'' as the occurrence of two past progenitors at a stellar mass greater than $1/5$.  The formation redshift, $z_{\rm form}$, was defined as the redshift at which the total mass of the halo was half of its mass at $z = 0$. We assumed this proxy for halo formation time in accordance with \cite{Mar20}. Although the latter authors defined $z_{\rm form}$ based on considerations encompassing ``centrals'' and  ``satellites'', whereas our analysis was focused on massive haloes, we adopted this definition for concreteness and as a reference for future work on ERPF in galaxy-sized halos.  In the present section, we only analysed the full TNG50-4 and TNG100-3 simulations, for {\tt Particle Type:} ``dm'', given the current availability of merger trees from {\tt SubLink}, which provides the mass value along the corresponding main progenitor branch of the {\tt Subfind} subhaloes at each snapshot \citep{Rod15}. 

In Fig. \ref{Pairplot_haloprops}, we present the paired density and scatterplot matrix for the above-mentioned variables (items {\it (i)} to {\it(iv)}), in order to isolate their potential correlations. We found that $\Delta M$ and $N_{\rm mer}$ were (individually) roughly correlated with the final total mass of the reference halo, but the correlation between the pair $\Delta M$ and $N_{\rm mer}$ seemed less evident. The variables $\Delta M$, $N_{\rm mer}$ and $z_{\rm form}$ roughly overlapped in their distribution ranges for both simulations. Most haloes in both simulations have assembled half of their final masses only at $z_{\rm form} \lesssim 2.0$. 

\begin{figure}
\centering
\includegraphics[trim= 0in 0in 0in 0in,clip,width=0.49\linewidth]{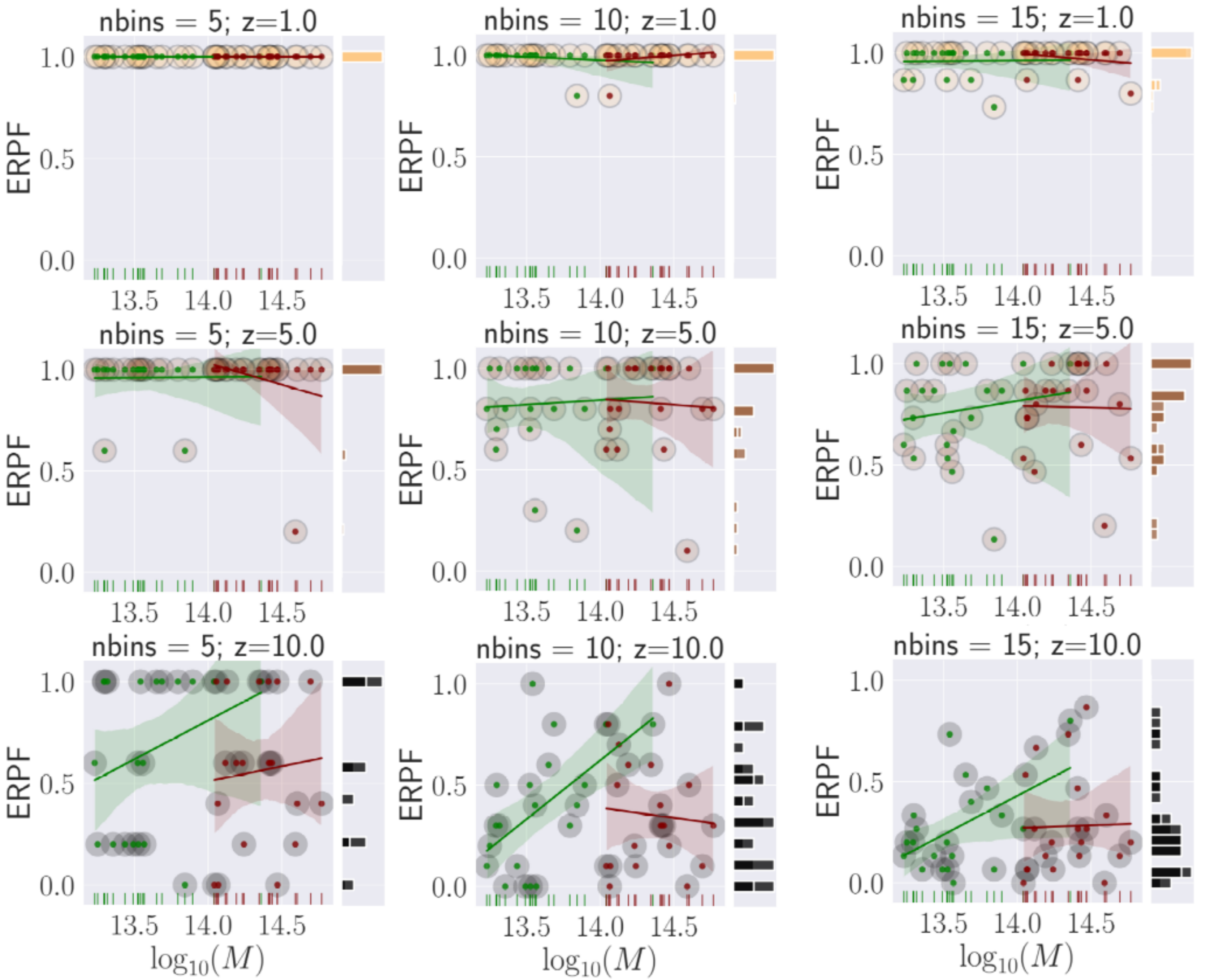} 
\includegraphics[trim= 0in 0in 0in 0in,clip,width=0.49\linewidth]{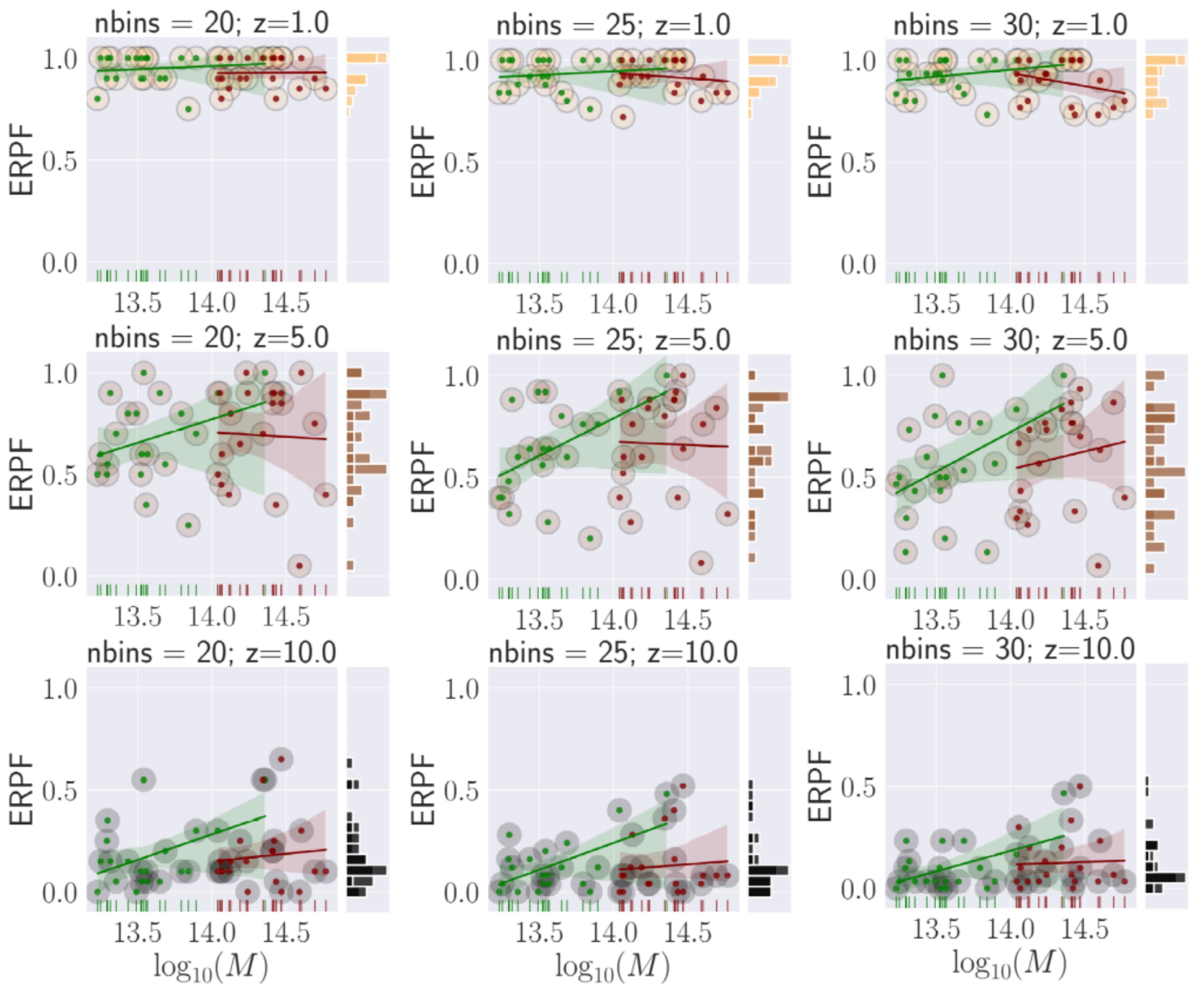} 
\caption{ERPF as a function of reference halo mass at $z=0$. Panels are distributed as a function of target redshift at which the ERPF was extracted, namely: top panels for {\tt z} $= 1.0$; middle, {\tt z} $= 5.0$; and bottom, {\tt z} $= 10.0$. From left to right, panels show increasing values of {\tt nbins}. In each panel, two sets of data points are included, namely: TNG5-40 full (lower mass range); and TNG100-3 full (higher mass range). To each set, a linear regression fit is provided, with a $95 \%$ confidence interval shown as translucent bands around each regression line. Histogram margins show the distribution of ERPF for all data in the given panel. \label{MVE}}
\end{figure}

\begin{figure}
\centering
\includegraphics[trim= 0in 0in 0in 0in,clip,width=0.49\linewidth]{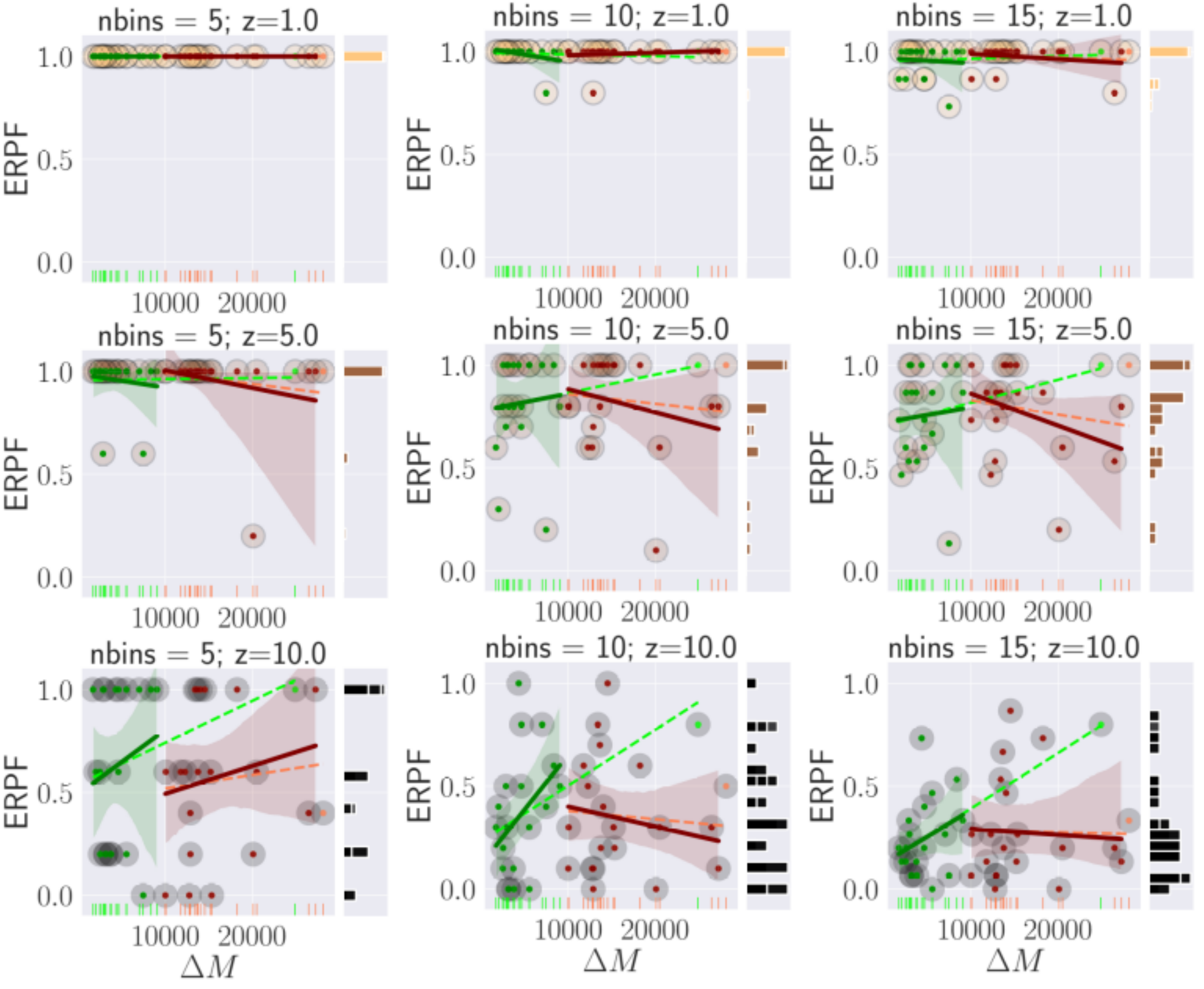} 
\includegraphics[trim= 0in 0in 0in 0in,clip,width=0.49\linewidth]{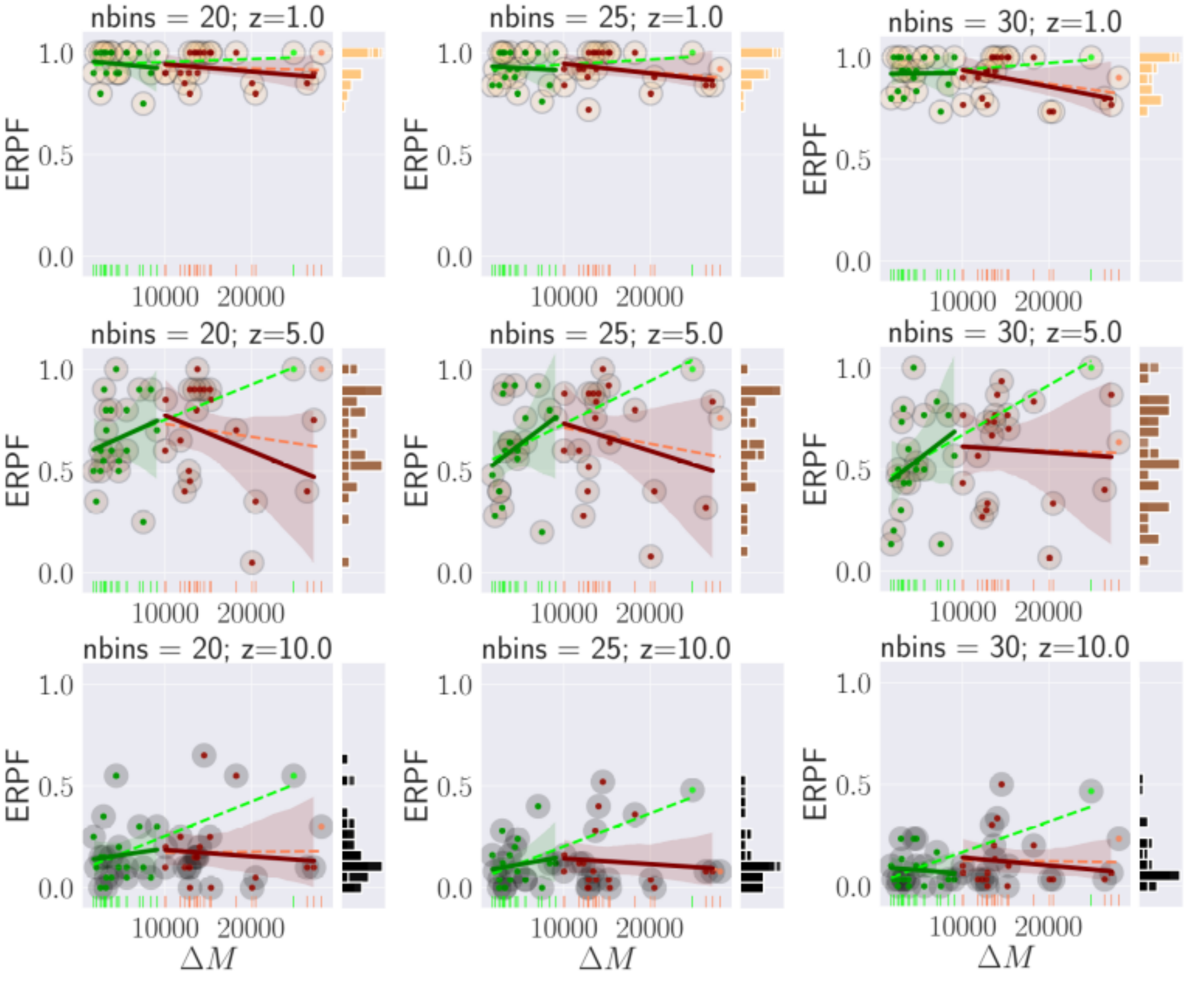} 
\caption{Same as the previous figure, but for ERPF as a function of the fractional increase in halo mass, $\Delta M$, using {\tt SubLink} data. The values of $\Delta M$ for the TNG100-3 (data points at right) were compressed by $50 \%$ and then shifted by a fixed value ($10^4$)  in order to avoid overlapping with the TNG50-4 data points (at left). Regression fits with (dashed line, lighter colour) and without (continuous line, darker colour) the highest $\Delta M$ point is provided. \label{DELTAM-ERPF}}
\end{figure}

\begin{figure}
\centering
\includegraphics[trim= 0in 0in 0in 0in,clip,width=0.49\linewidth]{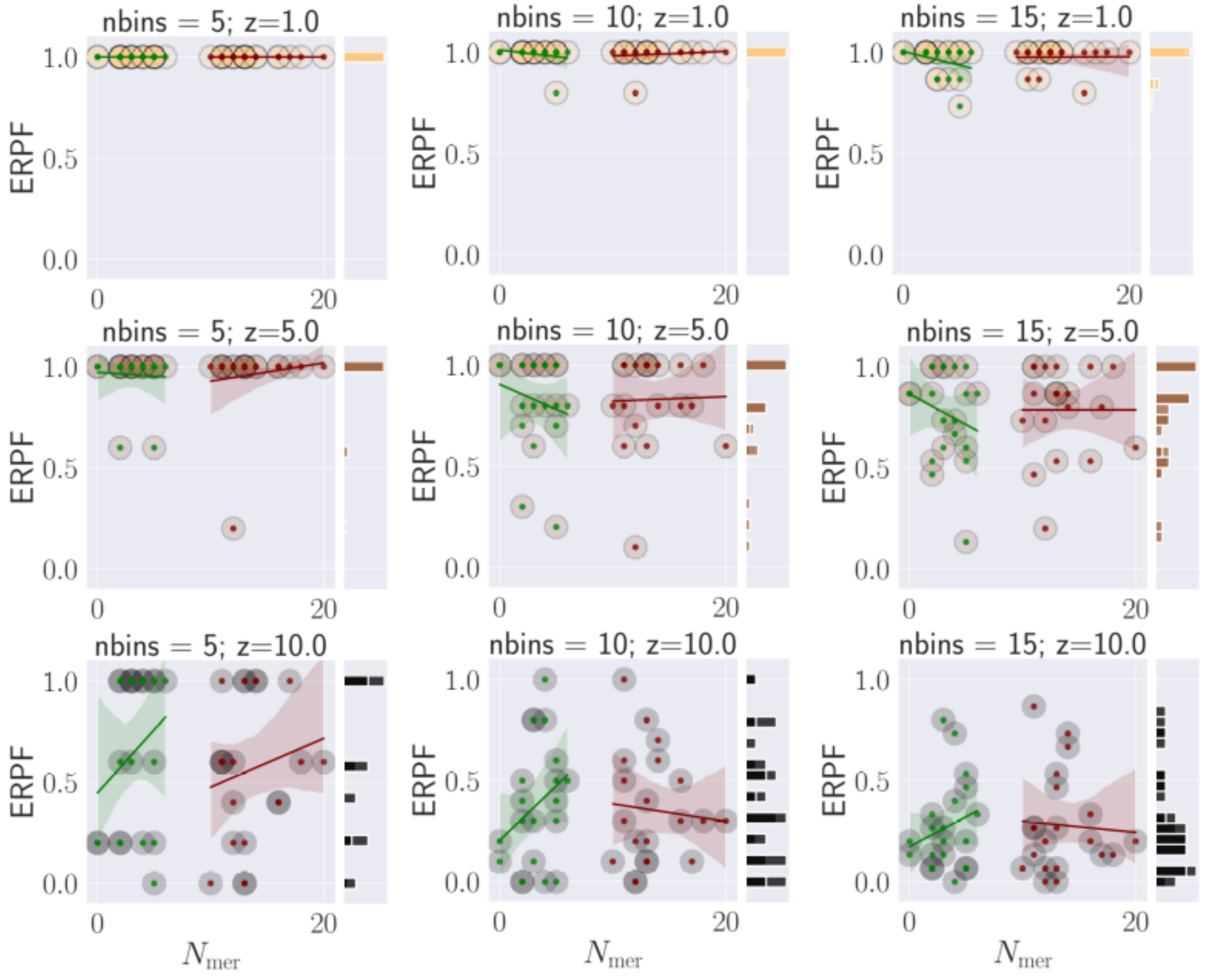} 
\includegraphics[trim= 0in 0in 0in 0in,clip,width=0.49\linewidth]{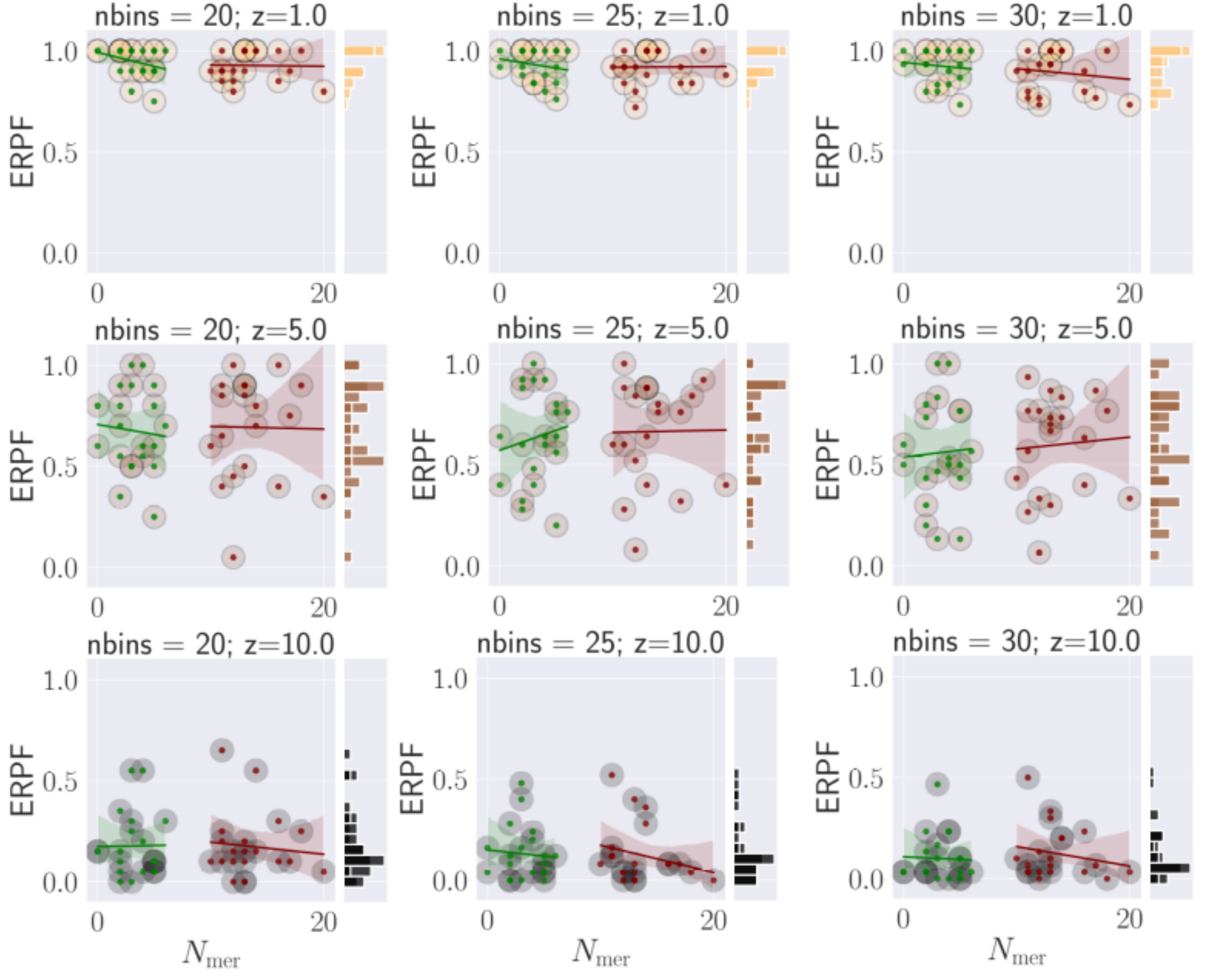} 
\caption{Same as the previous figure, but for ERPF as a function of the number of major mergers, $N_{\rm mer}$ (c.f. the main text for details). The values of $N_{\rm mer}$ for the TNG100-3 (data points at right) were shifted by a fixed value ($10$) in order to avoid overlapping with the TNG50-4 data points (at left). \label{NMERGERS-ERPF}}
\end{figure}

\begin{figure}
\centering
\includegraphics[trim= 0in 0in 0in 0in,clip,width=0.49\linewidth]{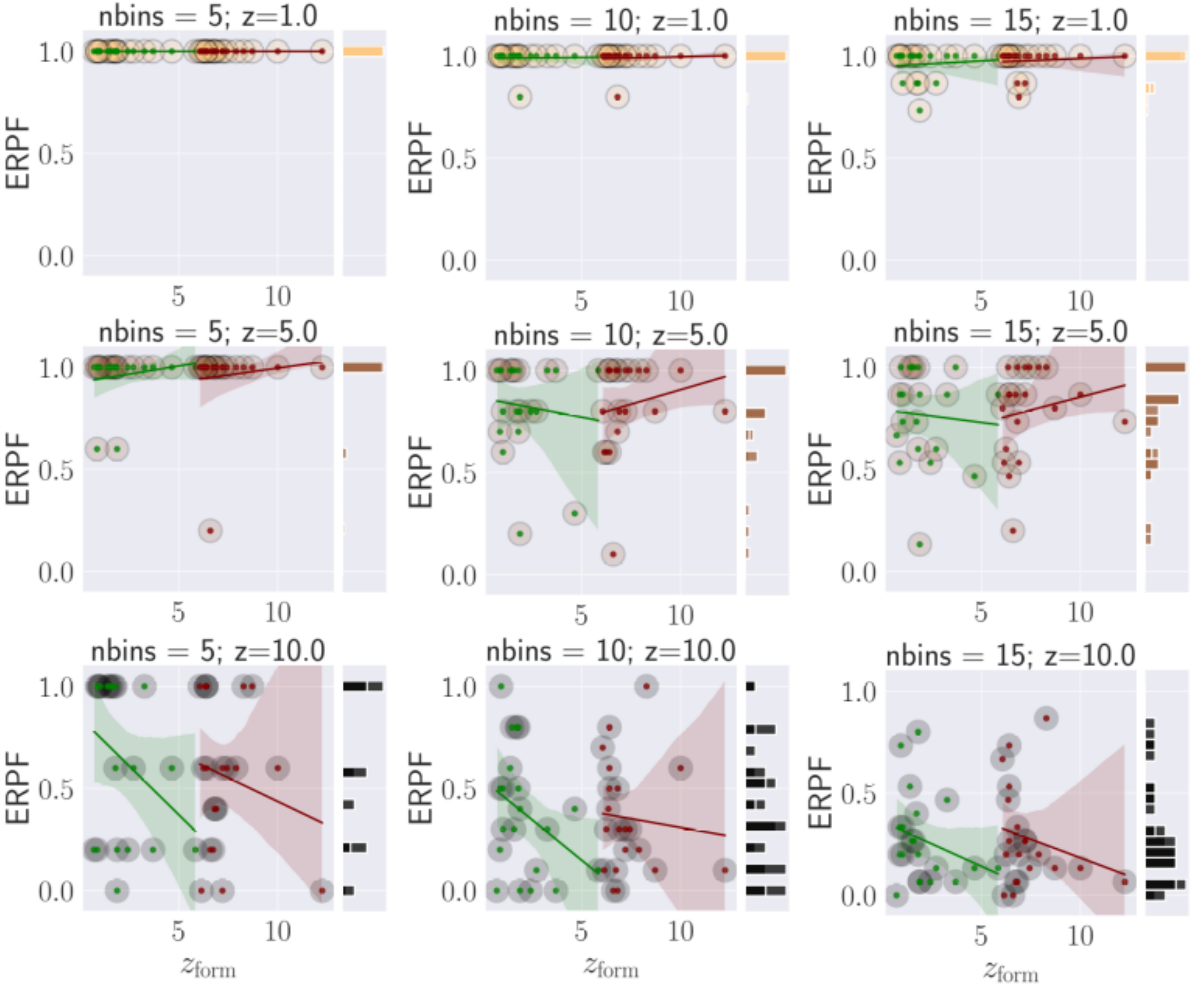} 
\includegraphics[trim= 0in 0in 0in 0in,clip,width=0.49\linewidth]{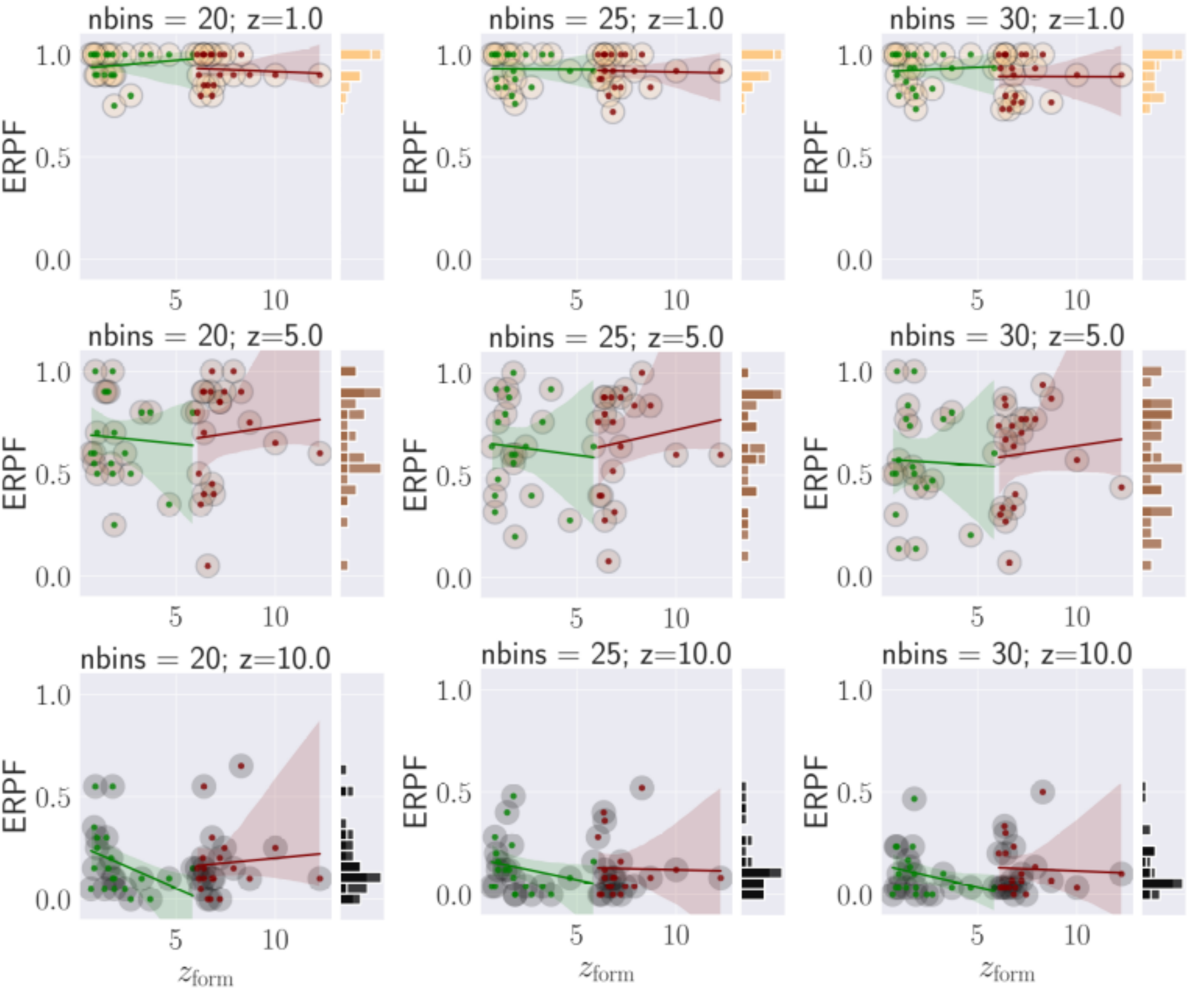} 
\caption{Same as the previous figure, but for ERPF as a function of the formation redshift, $z_{\rm form}$ (c.f. the main text for details). The values of $z_{\rm form}$ for the TNG100-3 (data points at right) were shifted by a fixed value ($5.0$) in order to avoid overlapping with the TNG50-4 data points (at left). \label{ZFORM-ERPF}}
\end{figure}

The behaviour of the ERP fractions as a function of the halo properties, {\it (i)} to {\it (iv)}, are displayed in Figs. \ref{MVE}, \ref{DELTAM-ERPF}, \ref{NMERGERS-ERPF}, and \ref{ZFORM-ERPF}, respectively  (i.e., total reference halo mass at $z=0$, $\Delta M$, $N_{\rm mer}$, and $z_{\rm form}$). These results are shown in separate panels in terms of reference redshift {\tt z} and {\tt nbins}. The common global trend seen in those figures was the expected decrease of ERPF as {\tt z} and {\tt nbins} increase. High ERPFs (above $\sim 0.8$) were a noticeable feature at {\tt z} $= 1.0$ for all {\tt nbins}, in both the TNG50-4 and TNG100-3 simulations.  

One important aspect of those panels is that not only they show how ERPFs might depend (or not) on halo properties and historical markers, but also how the {\it decline} in ERPFs proceeds as a function of the latter properties, as {\tt z} and {\tt nbins} increase. Indeed, the decline in ERPFs seemed to be a function of halo mass for the TNG50-4 case  (Fig. \ref{MVE}), for all {\tt nbins}. This trend was not seen in the haloes traced in the larger simulation box, TNG100-3: the ERPFs declined roughly in a ``uniform manner'', i.e., insensitively to halo mass. Indeed, in \cite{Dan06CEL} (c.f. their Fig. 2), the ERPFs for $31$ dark matter haloes (with masses less than $\sim 10^{14}$, at marker redshift {\tt z} $=10.0$; {\tt nbins} $=10$) resulted in more massive haloes showing more rank preservation than less massive ones (note that the $\theta$ parameter in that reference is the inverse of ERPF, that is, it measured the violation fraction).

For $\Delta M$ (Fig. \ref{DELTAM-ERPF}), the dependence of decline in ERPF on the increase of fractional mass persisted for TNG50-4, even after removing the apparent outlier (highest $\Delta M$ point), although the uncertainty on this trend was large. On the other hand, TNG100-3 seemed to show an opposite trend of decline in ERPF, for some intermediate panels, but essentially no trend at all in most of them (with large scatter). The situation was less clear in the case of $N_{\rm mer}$ (Fig. \ref{NMERGERS-ERPF}), or at least, the decline in ERPF seemed to be less sensitive to this variable. In the case of $z_{\rm form}$ (Fig. \ref{ZFORM-ERPF}), the scatter was also large, and at least for the $20$ analysed haloes, indicated independence of ERPFs decline with their ``formation'' redshift.

\section{Discussion and Conclusion}
\label{DIS}

ERP provides an alternative standpoint for analysing the problem of mixing in gravitational N-body systems.  ERP was proposed by \cite{Kan93}, almost $30$ years ago, to be indicative of some  mesoscopic constraint, operating at the level of collective particles in energy space. The physical nature of such a constraint, if real, remains unclear.  The purpose of this work was to verify whether previous results in the literature related to this effect were also reproduced in cosmological simulations with magnetohydrodynamical evolution and semi-analytical modelling (IllustrisTNG). For the first time, a  comparison between the ERP levels in the dark-only and full runs, i.e., in the absence and presence of the baryonic component, respectively, was obtained, as well as an analysis for ``star'' particles only, as they dynamically mixed within dark matter haloes. We again point out that the TNG50 and TNG100 simulations differ in their top $20$ massive haloes, so we probed different ranges in mass. Therefore, any comparisons between both simulations should be understood in terms of different mass ranges being probed, and not as a one-to-one comparison between the most massive haloes in those simulations. Our purpose was to advance the quantitative characterisation of ERP, prompted by open questions raised in \cite{Kan93} and by previous hints that ERP behaves in a complex manner. We summarise our results as follows:

{\it 1. Main overall ERP characterisation.} At the redshift marker {\tt z} $= 1$, we found high ERP fractions (above $\sim 80 \%$) in both simulations, for all {\tt nbins}. This result extended, in some cases, to ``star'' particles (although the statistics was significantly poorer for this baryonic component). Our results confirmed earlier indications in the literature of a significant level of  ``collective memory'' preservation of particle energies in the evolution of dark matter haloes, leading to a possible mesoscopic constraint {\it operative in a time span of several Gyr}.

{\it 2. Collective range for ERPF effectiveness.}  ERPFs in the analysed simulated haloes gradually declined for finer partitions (higher {\tt nbins}), as naturally expected. However, it is interesting to note that, in reference to the toy models of Sec. \ref{TOY}, we found that tightly distributed (i.e., less spread) particles in the one-particle energy space lead to a remarkable {\it consistency} of EPRFs, regardless of the choice of {\tt nbins}. This means that, for particles that evolve in a sufficiently {\it compact} manner in the energy space, ERPFs do not depend on the chosen coarse-graining level: they somewhat faithfully indicate the fractions of energy bin rank preservations for whatever {\tt nbins} used. 

{\it 3. Evolutionary range for ERPF effectiveness.}  ERPFs gradually declined for higher redshift markers, as expected. However, the manner in which this occurred was found to be, roughly, a function of halo mass (Fig. \ref{MVE}), at least  for the TNG50-4 case. This trend was not seen in the haloes traced in the larger simulation box, TNG100-3. Indeed, when tracing the decline of ERPF as a function of fractional increase in halo mass, this trend seemed to persist for TNG50-4 (Fig. \ref{DELTAM-ERPF}). This maybe partially due to a correlation between both variables (Fig. \ref{Pairplot_haloprops}), although the uncertainty seemed greater in the decline of ERPF with $\Delta M$.  In the case of the behaviour of ERPF in terms of the number of major mergers, or formation redshift, the decline of ERPF did not seem to be sensitive to these parameters, at least for the $20$ analysed haloes (Figs. \ref{NMERGERS-ERPF}, \ref{ZFORM-ERPF}). In any case, it is interesting to note that the decline in ERPF proceeded in {\tt z} in a relatively orderly sense.

{\it 4. ERPF effectiveness and dissipation.} Our analysis extended previous results in the literature, which have only been obtained for dark matter  simulations. We found that ERP could also be found in collections of stars, despite their dissipational histories, at least back from redshift marker {\tt z} $= 1$.  Approximately, the most (less) gravitationally bounded masses today were probably the most (less) bounded ones at redshifts even as high as $z \sim 5$, in some cases, regardless of astrophysically relevant dissipative mechanisms and efficient mixing in phase space. A comparison of of Dark vs. full (``dm'') halo pairs showed that the EPRFs were qualitatively similar, indicating that dissipative processes did not play a role in modifying the ERPFs of the dark matter component. However, clear exceptions were found, which could be an indication of some feedback process by the baryonic component, but this effect could also be entangled with purely statistical variations, specially for higher {\tt nbins} and {\tt z}. The relatively high level of ERPFs  for ``star''  particles (baryons), in some cases, at least back from $z = 1$, is a result that is also compatible with that of Wang et al., in which they found that a significant fraction of baryons in haloes (corresponding to the Milky Way scale) was accreted diffusively, rather than in major mergers, being relatively preserved at larger distances from the centre of the halo. This suggests that a further investigation of ERP in relation to the mass growth of galaxies through major and minor mergers in cluster of galaxies could provide an independent measure of environment effects on the evolution of such galaxies \citep[][and references therein]{Dre80,Rhe17}.

{\it 5. ERPF effectiveness, gravitational potential, and assembly history.}  The occurrence of significant ERPFs and their decline in time as a function of mass, Fig. \ref{MVE} (at least in the case of halo masses smaller than $\sim 10^{14} M_{\odot}$; see also previous indications in \citealt{Dan06CEL}) could be associated with an independent assessment of their assembly history. That is, our results are consistent with other studies indicating that dark matter haloes form in an inside-out fashion \citep[e.g.,][and references therein]{Wan11}, an effect verified from a different standpoint. By analysing (dark matter only) N-body simulations of six haloes with masses compatible with  those hosting galaxies at about the Milky Way scale, Wang et al. found that dark matter particles belonging to central regions (i.e., with the most negative binding energies) were already relatively stable in those regions at high redshifts. Such particles, belonging to the bottom of the halo potential well, could only be disturbed by major mergers, but the latter were more common at  high redshifts and less common at recent times. On the other hand, in the less bounded, external regions of haloes, particles were mostly gathered through a ``gentler''  process (diffusely or in minor mergers), generally keeping lower binding energies across several redshifts, until recent times. Hence, the overall process just described in terms of the configuration space of particles (i.e., roughly, a radial assembly gradient in term of accretion time) is a counterpart effect of ERP in the energy space. 

{\it 6. ERPF effectiveness and merger events.}  As already noted, the decline of EPPF with redshift (Fig. \ref{NMERGERS-ERPF}) did not seem to depend on the number of major mergers, but possibly on the fractional increase in mass (Fig. \ref{DELTAM-ERPF}), which evidently must include minor mergers. However, this dependence reveals a complex situation, as there seems to be a different trend for smaller (TNG50-4) and larger (TNG100-3) halo masses. An analysis on ERPF (or its decline) and the ``formation'' redshift (when the halo reached half of its final mass) did not indicate clear correlations (c.f. Fig. \ref{ZFORM-ERPF}). In any case, one would expect that a high ERPF up to a given redshift would be an imprint of a low fractional mass accretion or few merger events since that redshift. Indeed, ERP seemed to be more frequently violated (even for coarsely-grained energy bins) in lower mass haloes, indicating their relative susceptibility to merger events affecting their inner structures. Further investigations of ERPF on galaxy-sized haloes, with increased time and mass resolutions, correlating with other structural parameters, are left for a future work.

{\it 7. ERPF effectiveness and cosmological evolution.} It is important to note that the original results about ERP \citep{Kan93} were obtained in a non-cosmological setting, so the first indications about this effect did not include an expanding background and/or frequent ambient effects. A limited hierarchical scheme in \cite{Dan03} indicated some nuances on the effectiveness of ERP when comparing mergers and collapses. Other investigations probed ERP effects in cosmological halos (e.g., \citealt{Dan06MNRAS}, \citealt{Dan06CEL}), finding hints of correlations between ERP and the dispersion of the pairwise velocity distribution, as well as mass. Our results indicate that {\it ERP is not only roughly present in a ``single event scenario'' of an isolated merger or a collapse, but also along an involved dynamics that is a function of external disturbances in time}. The fact that such a complex dynamics can be correlated with the assembly time of haloes is an indication that violent relaxation \citep{Lyn67} is not complete, as has been discussed in the literature \citep[see mini-review in, e.g.,][and references therein]{Dan06MNRAS}. Diffuse accretion, major and minor mergers, infall, ejection of particles, etc, seem to be historically imprinted as a radial, coarse-grained gradient of accreted material in the structure of haloes, which belong to a collectively active environment. Hence, whatever the nature of an energy mesoscopic constraint, already operative in isolated events on a non-expanding background, {\it it seems to compound in the same direction as the hierarchical building up of haloes through cosmological times}.

We conclude with a rough picture in which the building up of dark matter haloes in a cosmological setting is a collective process in which particles closely spaced in coarse-grained cells in energy space can efficiently mix in the phase space corresponding to those energy cells. The hierarchical growth of haloes produce structural correlations as they settle in roughly stratified equilibrium configurations as a function of local relaxation timescales after external disturbances and accretion. However, {\it there seems to be an orderly restriction to the exchange of energy between those energy cells}, so that the latter represent elements moving  in energy space which are roughly preserved. That is, their energy boundaries can only change linearly and orderly with respect to all other energy cells of the system. A detailed theory of this linear and rank preserving process, at the level of the coarse-grained energy distribution function, is yet to be formally devised. The present work hopefully advances a preliminary map of the parameter space required for such a development.


\section*{Acknowledgements}

We thank the anonymous referee for suggestions that improved our paper.
This study was financed in part by the Coordena\c c\~ao de Aperfei\c coamento de Pessoal de N\'{\i}vel Superior - Brasil (CAPES) - Finance Code 001, the Programa Institucional de Internacionaliza\c c\~ao (PrInt – CAPES), and the Brazilian Space Agency (AEB) for the funding (PO 20VB.0009).



\section*{DATA AVAILABILITY STATEMENT}

The data underlying this article will be shared on reasonable request to the corresponding author.



\bibliographystyle{mnras}
\bibliography{CCDANTAS}


\appendix

\section{Energy ranking fractions for remaining analysed data}
\label{APP1}

\begin{figure}
\centering
\includegraphics[trim= 0in 0in 0in 0in,clip,width=0.8\linewidth]{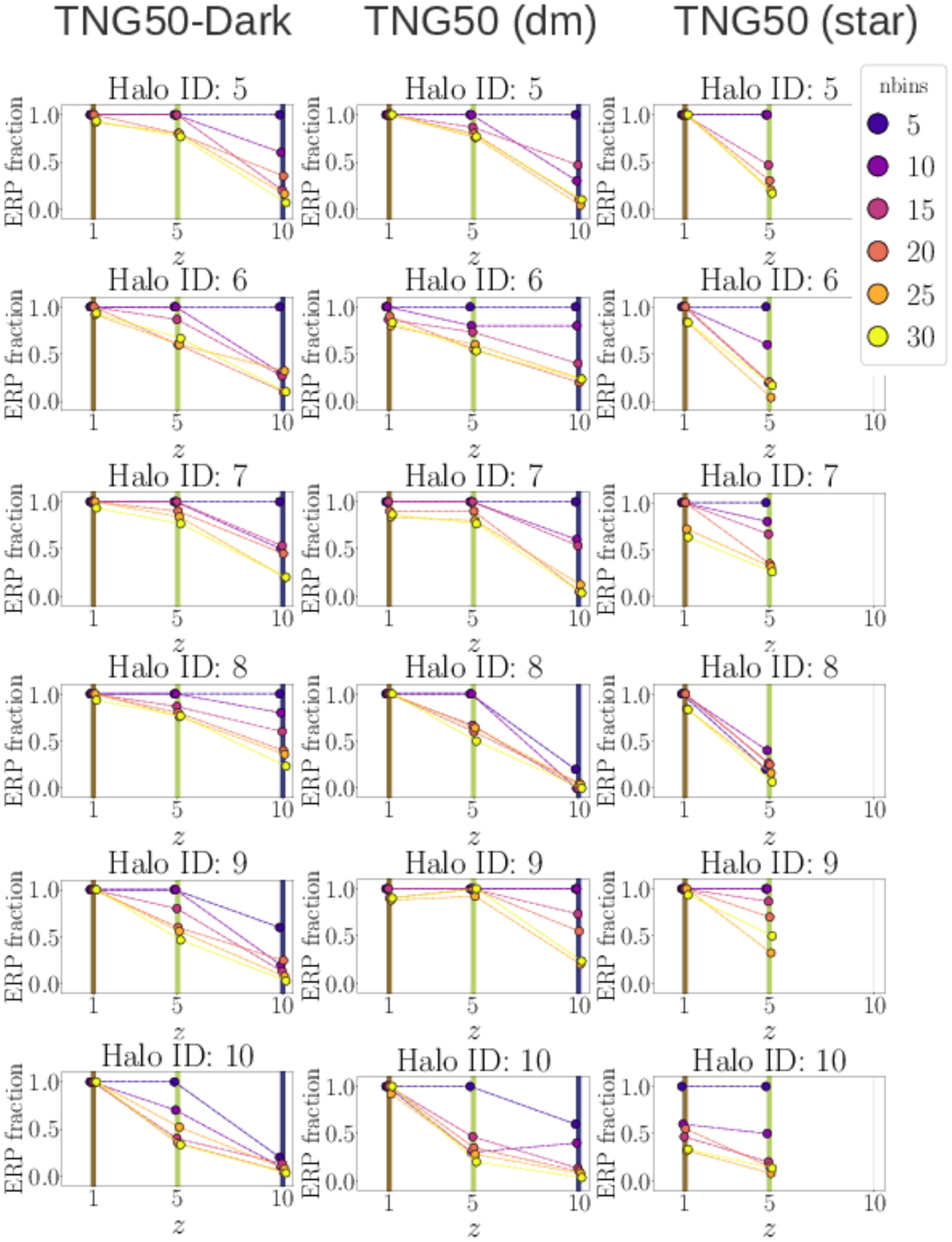} 
\caption{ERP fractions for {\tt Halo ID} $= \{ 5, \dots, 10 \}$, for TNG50-4-Dark and TNG50-4 ({\tt Particle Type:} ``dm'' and ``star'', respectively).  \label{TNG50-5-10}}
\end{figure}

\begin{figure}
\centering
\includegraphics[trim= 0in 2in 0in 0in,clip,width=0.8\linewidth]{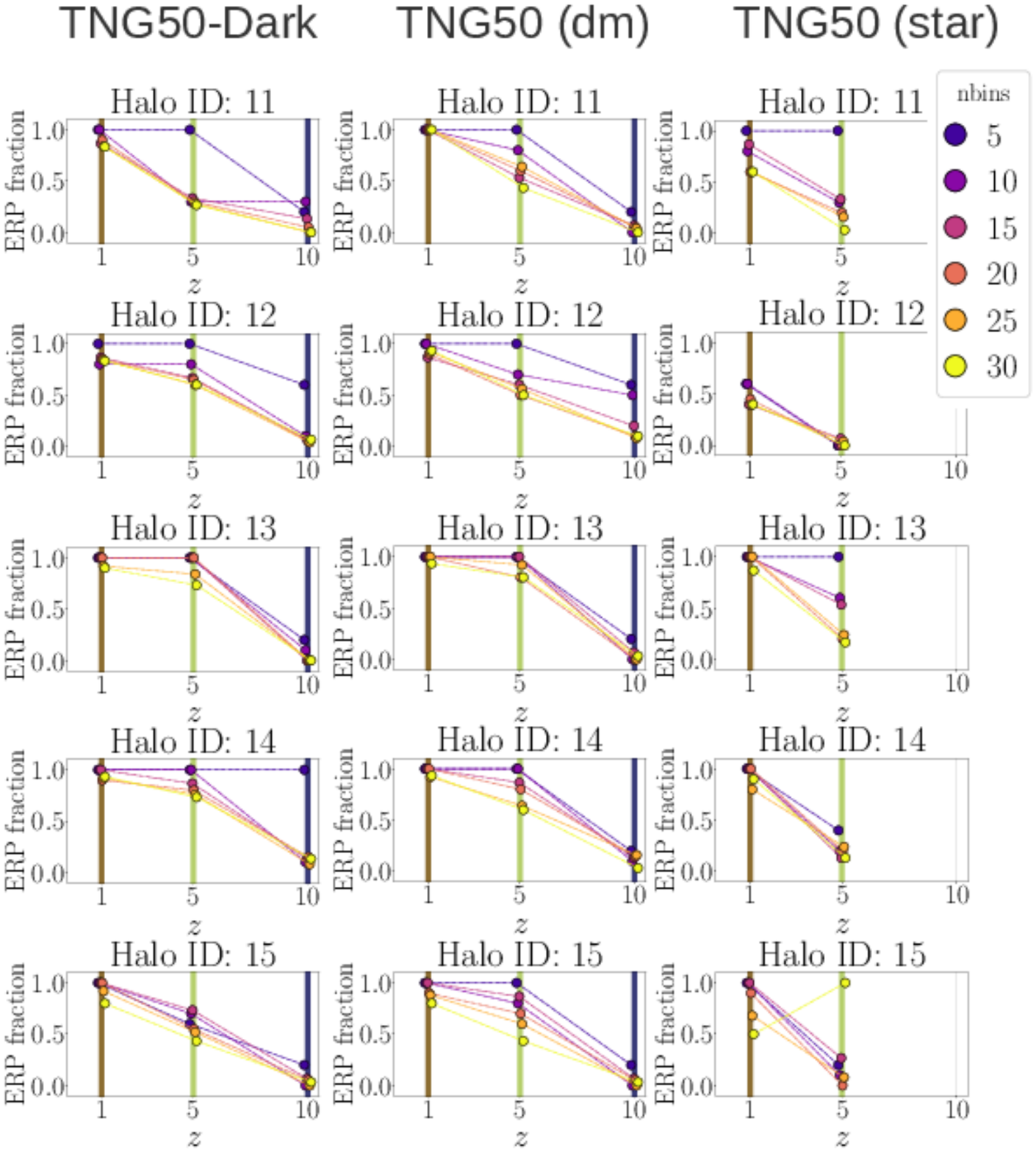}
\caption{ERP fractions for {\tt Halo ID} $= \{ 11, \dots, 15 \}$, for TNG50-4-Dark and TNG50-4 ({\tt Particle Type:} ``dm'' and ``star'', respectively).  \label{TNG50-11-15}}
\end{figure}

\begin{figure}
\centering
\includegraphics[trim= 0in 2in 0in 0in,clip,width=0.8\linewidth]{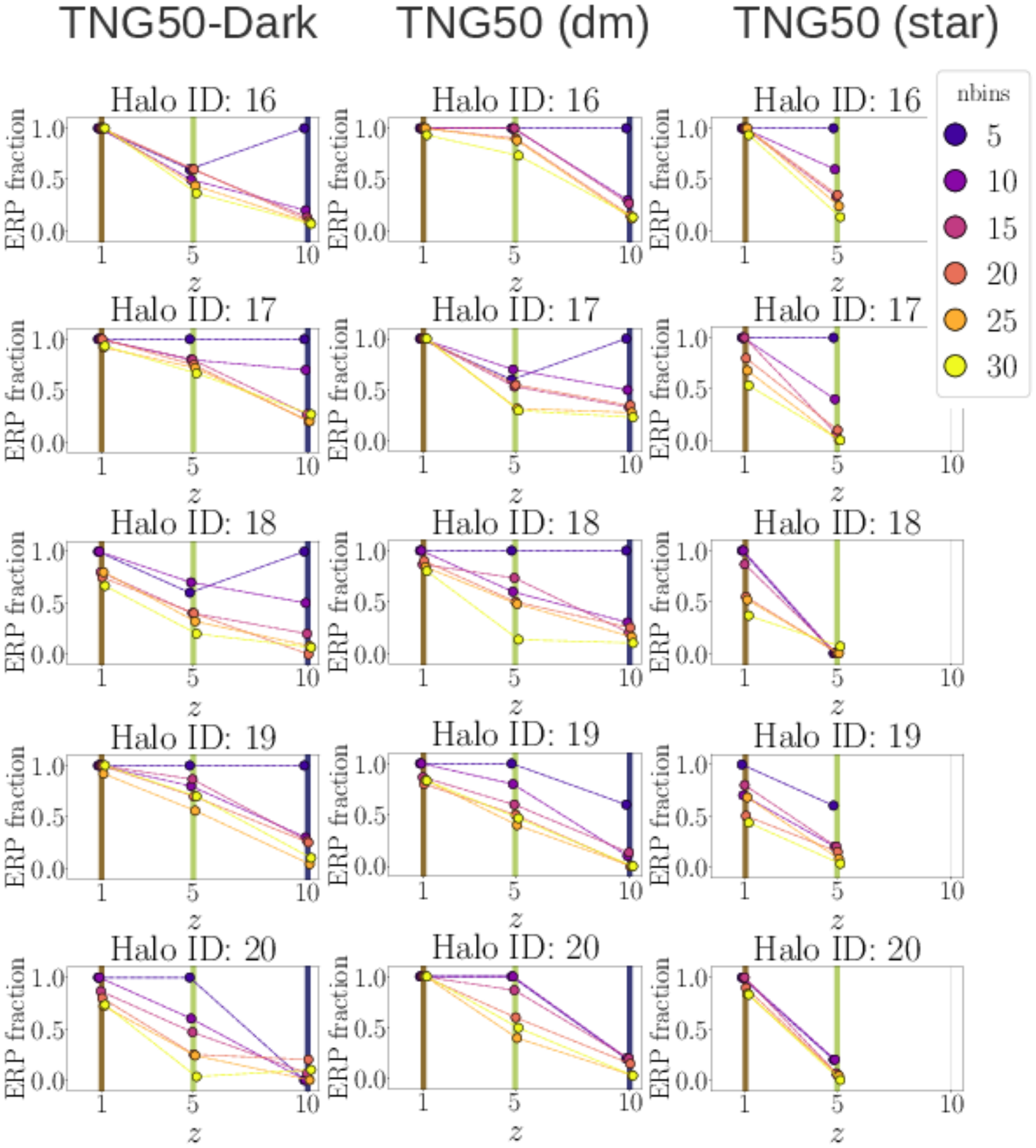}  
\caption{ERP fractions for {\tt Halo ID} $= \{ 16, \dots, 20 \}$, for TNG50-4-Dark and TNG50-4 ({\tt Particle Type:} ``dm'' and ``star'', respectively).  \label{TNG50-16-20}}
\end{figure}

\begin{figure}
\centering
\includegraphics[trim= 0in 0in 0in 0in,clip,width=0.8\linewidth]{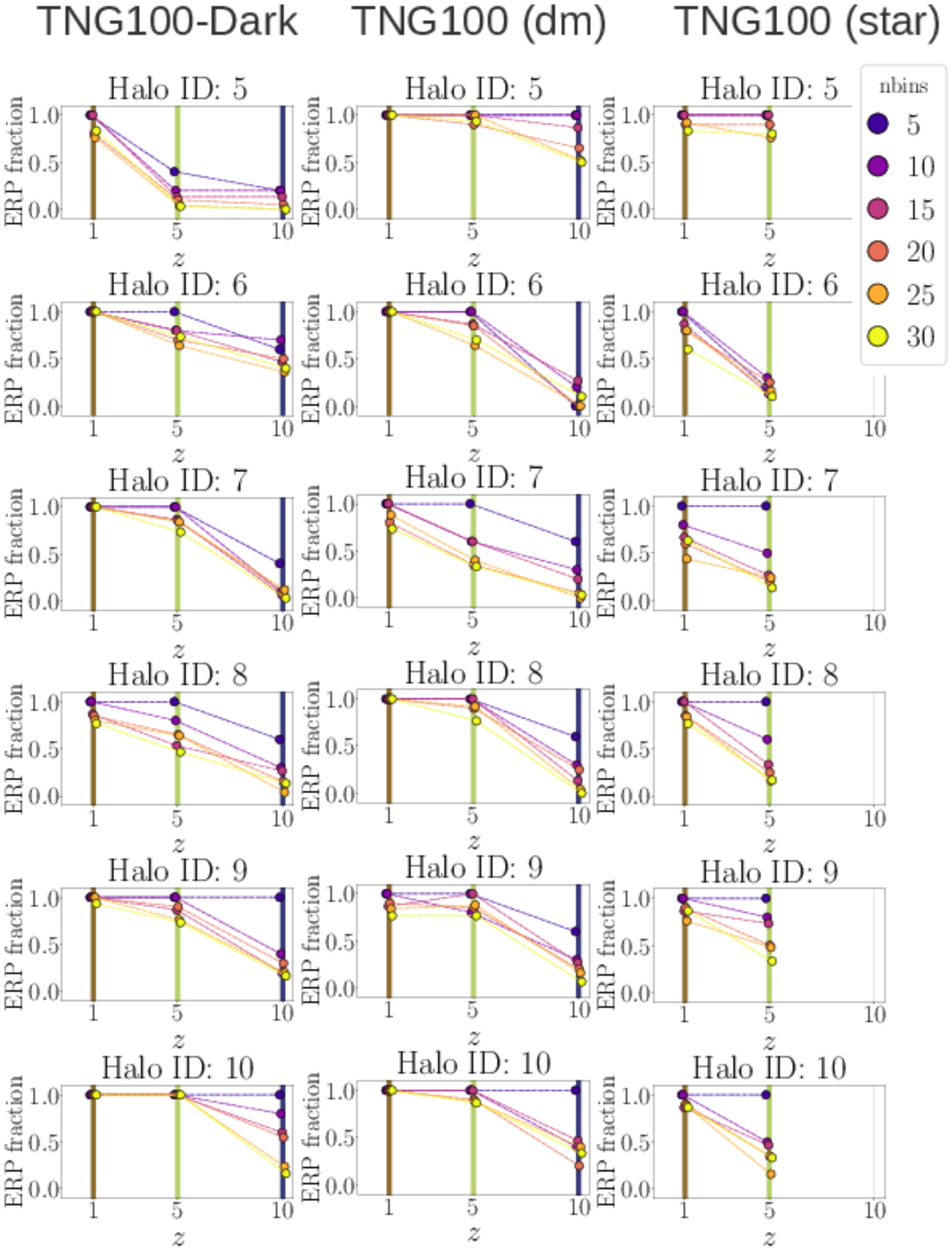}  
\caption{ERP fractions for {\tt Halo ID} $= \{ 5, \dots, 10 \}$, for TNG100-3-Dark and TNG100-3 ({\tt Particle Type:} ``dm'' and ``star'', respectively).  \label{TNG100-5-10}}
\end{figure}

\begin{figure}
\centering
\includegraphics[trim= 0in 2in 0in 0in,clip,width=0.8\linewidth]{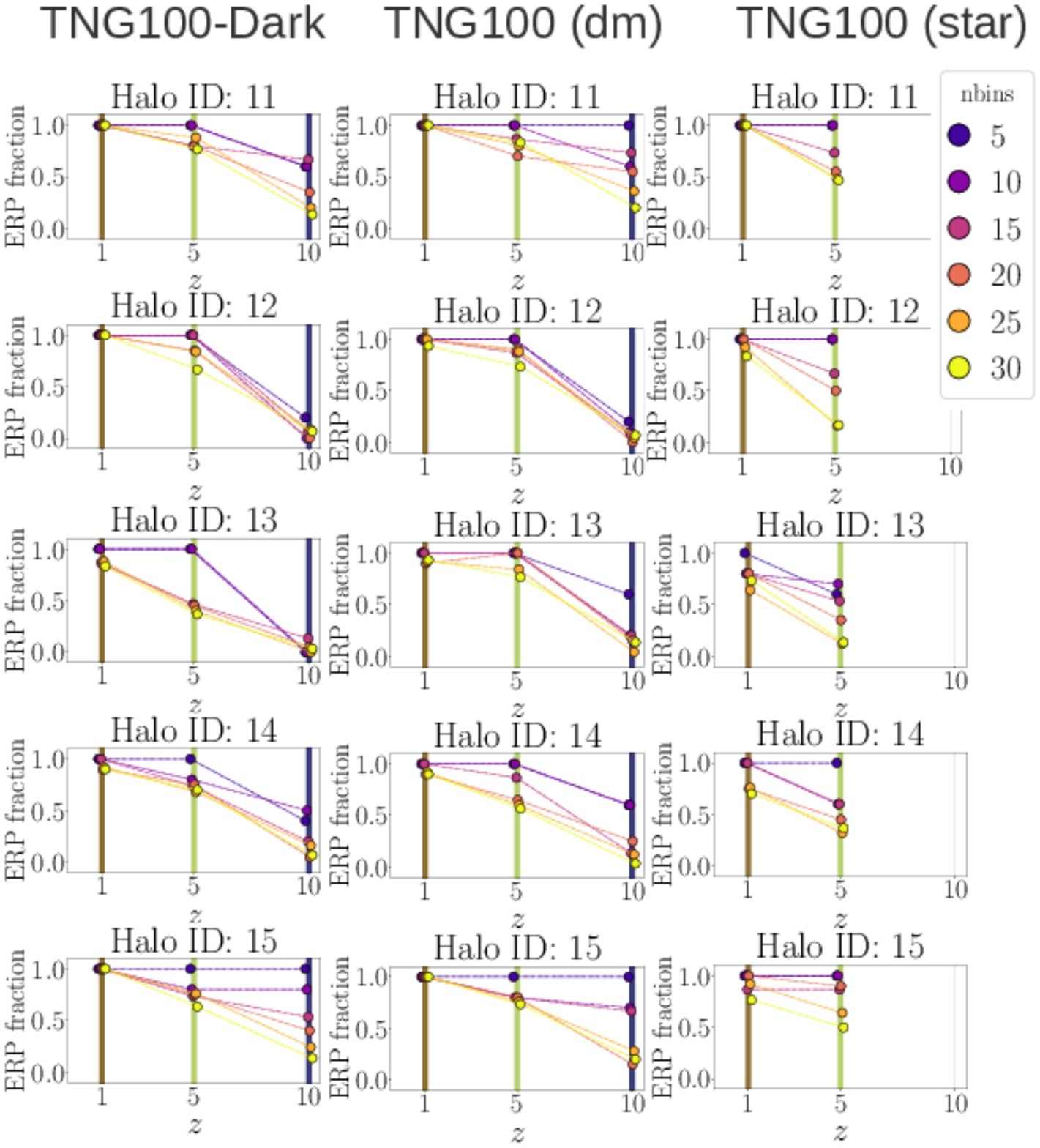} 
\caption{ERP fractions for {\tt Halo ID} $= \{ 11, \dots, 15 \}$, for TNG100-3-Dark and TNG100-3 ({\tt Particle Type:} ``dm'' and ``star'', respectively).  \label{TNG100-11-15}}
\end{figure}

\begin{figure}
\centering
\includegraphics[trim= 0in 2in 0in 0in,clip,width=0.8\linewidth]{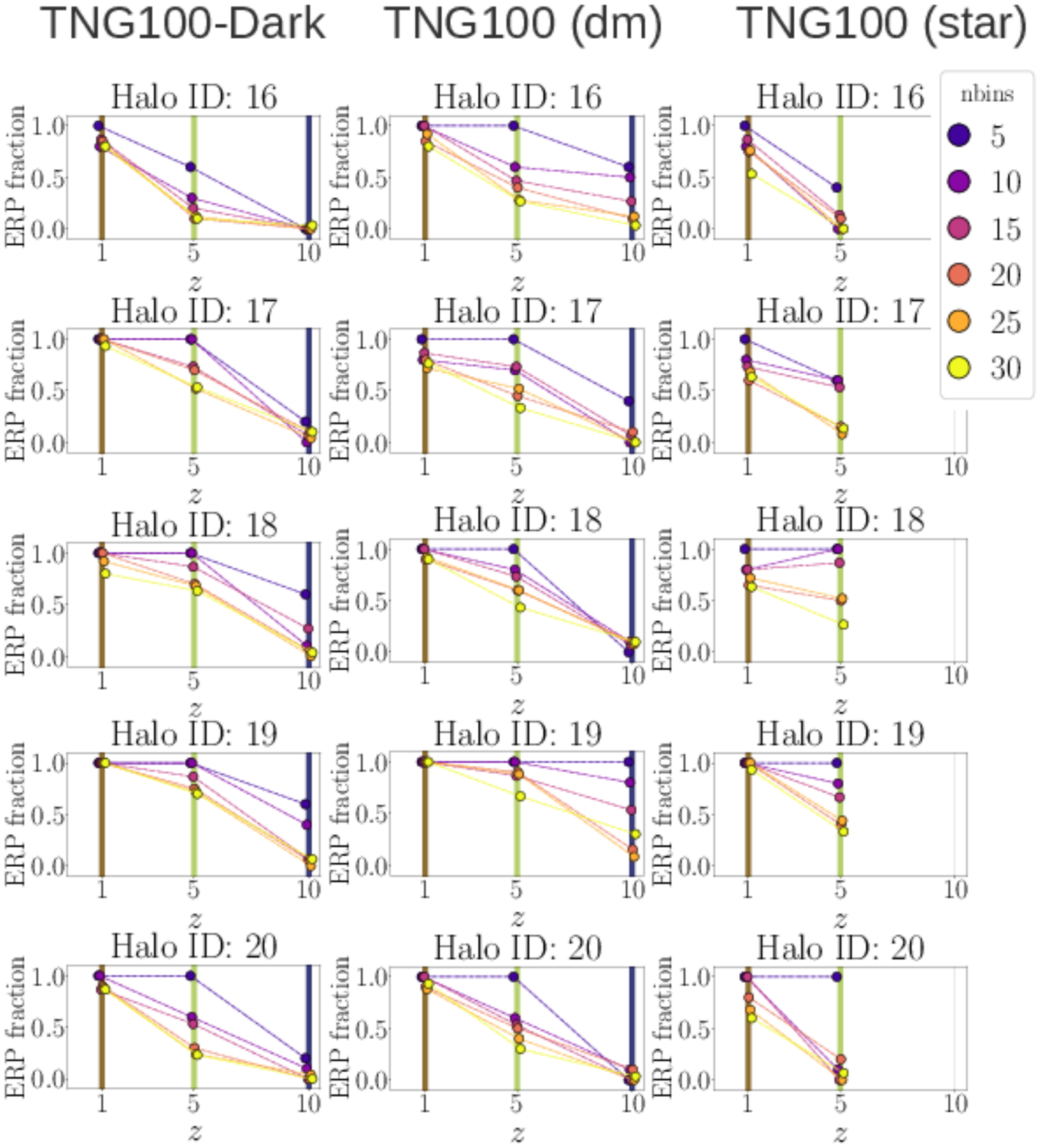}  
\caption{ERP fractions for {\tt Halo ID} $= \{ 16, \dots, 20 \}$, for TNG100-3-Dark and TNG100-3 ({\tt Particle Type:} ``dm'' and ``star'', respectively).  \label{TNG100-16-20}}
\end{figure}

We present in Figs. \ref{TNG50-5-10},  \ref{TNG50-11-15},  \ref{TNG50-16-20} (TNG50) and Figs. \ref{TNG100-5-10},  \ref{TNG100-11-15},  \ref{TNG100-16-20} (TNG100) the remainder ERP fractions, i. e., for {\tt Halo ID} $= \{ 5, \dots, 20 \}$.

\bsp	
\label{lastpage}

\end{document}